\documentclass[aps,prb,amsmath,amssymb,reprint,superscriptaddress]{revtex4-2}

\usepackage{lineno,hyperref}
\usepackage{graphicx}

\begin{document}

\title{\texorpdfstring{Electrical detection of the spin reorientation transition in antiferromagnetic TmFeO$_3$ thin films by spin Hall magnetoresistance}{Electrical detection of the spin reorientation transition in antiferromagnetic TmFeO3 thin films by spin Hall magnetoresistance}}

\author{S.~Becker}
	\email{svenbecker@uni-mainz.de}
\affiliation{Institute of Physics, University of Mainz, Staudingerweg 7, 55128 Mainz}
\author{A.~Ross}
\affiliation{Institute of Physics, University of Mainz, Staudingerweg 7, 55128 Mainz}
\affiliation{Graduate School of Excellence “Materials Science in Mainz” (MAINZ), Staudingerweg 9, 55128 Mainz}
\author{R.~Lebrun}
\affiliation{Institute of Physics, University of Mainz, Staudingerweg 7, 55128 Mainz}
\affiliation{Unité Mixte de Physique CNRS, Thales, University Paris-Sud, Université Paris-Saclay, Palaiseau 91767, France}
\author{L.~Baldrati}
\affiliation{Institute of Physics, University of Mainz, Staudingerweg 7, 55128 Mainz}
\author{S.~Ding}
\affiliation{Institute of Physics, University of Mainz, Staudingerweg 7, 55128 Mainz}
\affiliation{Graduate School of Excellence “Materials Science in Mainz” (MAINZ), Staudingerweg 9, 55128 Mainz}
\affiliation{State Key Laboratory for Mesoscopic Physics, School of Physics, Peking University, Beijing 100871, China}
\author{F.~Schreiber}
\affiliation{Institute of Physics, University of Mainz, Staudingerweg 7, 55128 Mainz}
\author{F.~Maccherozzi}
\author{D.~Backes}
\affiliation{Diamond Light Source, Harwell Science and Innovation Campus, Didcot OX11 0DE, United Kingdom}
\author{M.~Kl\"aui}
\affiliation{Institute of Physics, University of Mainz, Staudingerweg 7, 55128 Mainz}
\affiliation{Graduate School of Excellence “Materials Science in Mainz” (MAINZ), Staudingerweg 9, 55128 Mainz}
\affiliation{Center for Quantum Spintronics, Norwegian University of Science and Technology, 7491 Trondheim, Norway}
\author{G.~Jakob}
\affiliation{Institute of Physics, University of Mainz, Staudingerweg 7, 55128 Mainz}
\affiliation{Graduate School of Excellence “Materials Science in Mainz” (MAINZ), Staudingerweg 9, 55128 Mainz}

\date{\today}
\begin{abstract}
TmFeO$_3$ (TFO) is a canted antiferromagnet that undergoes a spin reorientation transition (SRT) with temperature between 82\,K and 94\,K in single crystals. In this temperature region, the N\'eel vector continuously rotates from the crystallographic $c$-axis (below 82\,K) to the $a$-axis (above 94\,K). The SRT allows for a temperature control of distinct antiferromagnetic states without the need for a magnetic field, making it apt for applications working at THz frequencies. For device applications, thin films of TFO are required as well as an electrical technique for reading out the magnetic state. Here we demonstrate that orthorhombic TFO thin films can be grown by pulsed laser deposition and the detection of the SRT in TFO thin films can be accessed by making use of the all electrical spin Hall magnetoresistance (SMR), in good agreement for the temperature range where the SRT occurs. Our results demonstrate that one can electrically detect the SRT in insulators.
\end{abstract}
\maketitle

\section{Introduction}
Antiferromagnets (AFMs) are  a key focus of current spintronic research \cite{RevModPhys.90.015005,Jungwirth2016,Jungwirth2018}. Due to their fast spin dynamics and stability against external perturbations, AFMs are promising candidates for THz emitters \cite{Sulymenko2017,Stremoukhov2019} and next generation data storage devices \cite{Wadley2016}. The insulating canted antiferromagnet TmFeO$_3$ (TFO) has recently gathered significant interest because of the possibility to coherently switch its spin state by $90^\circ$ with a THz pulse \cite{Schlauderer2019} in the proximity of the temperature-driven spin reorientation transition (SRT) in bulk single crystals. 
The detection of the switching was achieved by the Faraday effect in a transmission geometry. However, not only are bulk crystals inappropriate for back-end integration into devices, but the requirements of writing and reading out by THz laser pulses necessitates large external power sources. Moving towards applications then requires high quality thin films possessing similar antiferromagnetic properties to their bulk counterparts and integrated mechanisms for controlling and reading the antiferromagnetic N\'eel vector. The growth of different antiferromagnetic thin films can be achieved via several techniques including sputtering \cite{Jourdan2015} and pulsed laser deposition \cite{Mix2013}. Meanwhile, current-induced control of the N\'eel vector has recently been shown in antiferromagnetic insulators \cite{NiOImaging}, whilst the electrical read out has been achieved making use of the spin Hall magnetoresistance (SMR) between a heavy metal and an antiferromagnetic insulator \cite{NiOImaging,Moriyama2018}. 
With respect to TFO thin films, growth has been achieved in the hexagonal phase when deposited on Al$_2$O$_3$ (0001), Pt (111) \cite{hTFO,hTFO2} and YSZ (111) \cite{TFOthin1} substrates. This hexagonal phase has different properties compared to the orthorhombic phase, like multiferroicity and low magnetic ordering temperature \cite{Akbashev2011}. Canted antiferromagnetism with the spin reorientation transition is so far only reported for the orthorhombic phase. Until now, there has only been one report of TFO thin films grown on SrTiO$_3$ (STO) substrates that describes the structure as the standard perovskite structure, without characterization of their magnetic properties \cite{TFOthin1}. It is then at present not clear if thin films present a similar spin structure and SRT as found in the bulk, but understanding this is a key requirement in view of applications.\\ 
Here, we demonstrate that TFO grows in fact in the bulk-like orthorhombic phase on STO, not in the cubic perovskite phase. Our thin films are of high crystallographic quality exhibiting oriented growth. We detect a canted moment and the spin reorientation transition similar to the bulk. Furthermore, we ascertain the direction of the N\'eel vector (\textbf{N}) and canted moment (\textbf{M}) in TFO thin films using SMR. We show how the spin reorientation can be read-out electrically, thus being promising in view of applications such as, as a source of monochromatic THz radiation or data storage devices.

\section{Results and Discussion}
\begin{figure}[t]%
\centering
	\includegraphics[width=1\columnwidth]{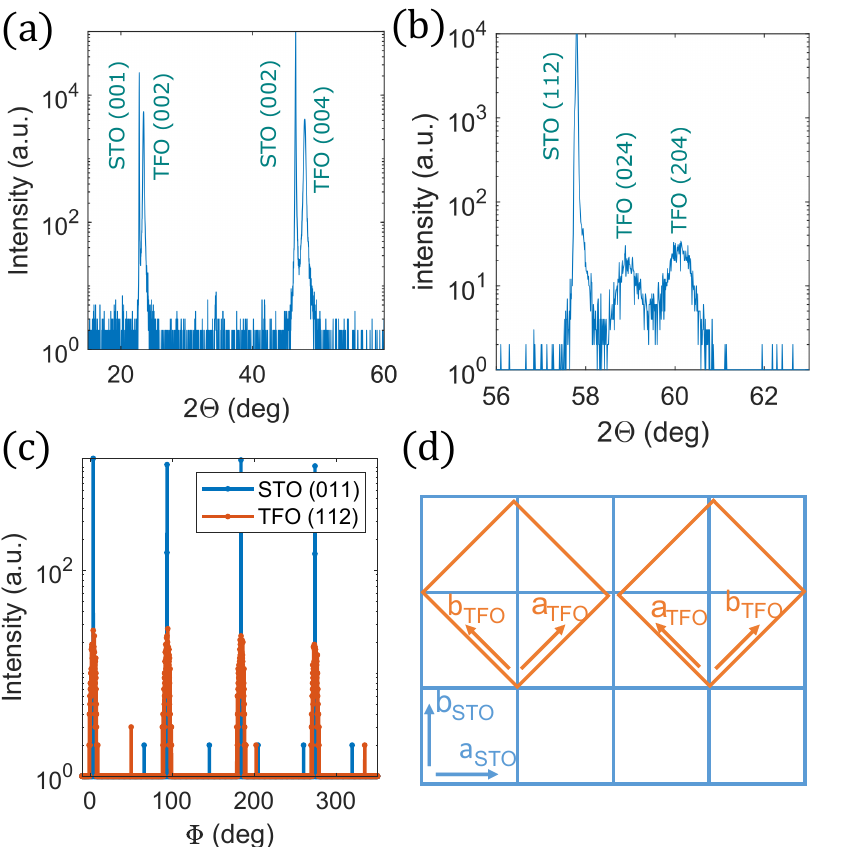}%
	\caption{(a) 2$\Theta/\omega$ scan of a 200\,nm TFO thin film grown on a STO (001)$_p$ substrate measured out-of-plane. (b) 2$\Theta/\omega$ measured along (112)$_p$ direction revealing both (024)$_o$ and (204)$_o$ TFO reflexes. (c) $\Phi$-scans with $2\Theta$ fixed at STO (011)$_p$ and TFO (112)$_o$ reflexes. (d) Relative orientation of the TFO unit cells (orange) with respect to the STO surface (blue). Both configurations are present across the sample surface. The cell dimensions are to scale showing a good fit of the TFO $b$ axis with the STO diagonal.}%
	\label{fig:XRD}%
\end{figure}

%
%
%
Using the deposition parameters detailed in the Methods section \ref{sec:Methods}, 200\,nm thick TFO$_o$ films were deposited on (001)$_p$ oriented STO substrates. The subscripts $o$ and $p$ indicate the directions in an orthorhombic and perovskite unit cell, respectively. The crystallographic structure was investigated by making use of x-ray diffraction (XRD). An out-of-plane XRD scan in the range of $2\Theta= 15^\circ - 60^\circ$ for a typical TFO sample is shown in FIG. \ref{fig:XRD} (a). 
We observe (001)$_p$ and (002)$_p$ substrate and (002)$_o$ and (004)$_o$ TFO peaks with no additional impurity phase present. We calculate an out-of-plane lattice constant of 7.58\,\AA, which corresponds to the value of the bulk orthorhombic TFO $c$-axis ($a=5.25$\,\AA, $b=5.57$\,\AA, $c=7.58$\,\AA) \cite{Marezio}. The other lattice parameters are determined by in-plane XRD measurements.
A 2$\Theta/\omega$ scan along the (112)$_p$ direction gives the diffraction pattern shown in FIG. \ref{fig:XRD} (b). Besides the (112)$_p$ STO substrate peak we find the (024)$_o$ TFO peak at $2\Theta=58.92^\circ$ and the (204)$_o$ TFO peak at $2\Theta=60.07^\circ$.
To calculate the lattice parameters we use the standard textbook formula for orthorhombic unit cells: $\frac{1}{d^2}=\frac{h^2}{a^2}+\frac{k^2}{b^2}+\frac{l^2}{c^2}$, where d is the spacing of diffraction planes calculated from the peak angle $h$, $k$ and $l$ Miller indices and $c$ the lattice constant calculated from the out-of-plane XRD scan. We find $a=5.28$\,\AA\ and $b=5.57$\,\AA. The presence of both (024)$_o$ and (204)$_o$ diffraction peaks in one $2\Theta/\omega$ scan leads to the conclusion that there are two crystallographic domains with different in-plane order present. The orthorhombic $b$-axis of TFO is along the $\left<110\right>_p$ of STO in one domain, along the $\left<\bar{1}10\right>_p$ in another. The domains are separated by a 90$^\circ$ in-plane rotation of the unit cells. This is confirmed by $\Phi$ scans as shown in FIG. \ref{fig:XRD} (c).
We find that the (112)$_o$ peaks in $\Phi$ direction have a FWHM of $5.6^\circ$ and reveal a four-fold in-plane symmetry not consistent with the two-fold symmetry expected from a single orthorhombic crystallographic domain. The (204)$_o$ peak on the other hand has a FWHM of $4.3^\circ$, also revealing a four-fold in-plane symmetry (not shown). The smaller width of the (204)$_o$ peaks compared to the (112)$_o$ peaks in the $\Phi$ scans indicate that the orthorhombic $a$ and $b$ axes align with the STO diagonals leading to a distribution of orientations of the orthorhombic diagonal. The relative orientations are schematically shown in FIG. \ref{fig:XRD} (d) for the unit cell of TFO (orange) atop the in-plane structure of the substrate (blue). Knowing the growth orientation of the TFO on STO we determine a lattice mismatch of $-4.9\%$ for the TFO $a$-axis and $+0.9\%$ for the TFO $b$-axis each corresponding to the diagonal of the STO $ab$-plane. The sketch is made to scale so one can see that the TFO $b$-axis fits well the STO diagonal. Our thin films clearly show the orthorhombic structure of bulk samples, which has not been reported before for thin films. \par
\begin{figure}[t]%
\centering
	\includegraphics[width=1\columnwidth]{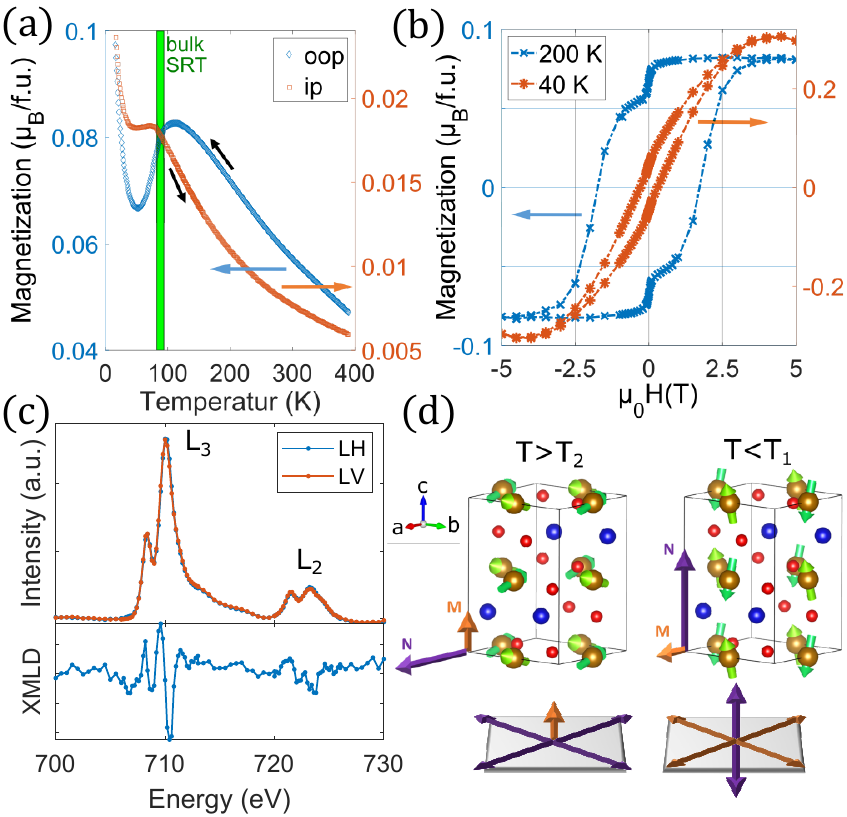}%
	\caption{(a) Magnetization vs temperature for out-of-plane (oop, along the $c$ axis) and in-plane (ip, along orthorhombic $a$ axis, $45^\circ$ towards STO axis) configuration. In green the bulk spin reorientation transition is depicted. (b) Field sweep with field applied in the oop direction along the $c$ axis with a maximum applied field of 5\,T. The data has been corrected for para- and diamagnetic background. (c) XAS spectra for linear horizontal (LH) and linear vertical (LV) polarization measured at 300\,K (top), XMLD spectrum (bottom) calculated as the difference between LV and LH. (d) The magnetic configuration of the TFO unit cell above (left) and below (right) the spin reorientation transition drawn in VESTA \cite{Vesta}. Due to the presence of two crystallographic domains the $a$-axis can be aligned in two directions leading to the magnetic configuration of the overall sample as sketched below corresponding to the cubic STO surface. In purple the N\'eel \textbf{N} vector, in orange the canted moment \textbf{M}.}%
	\label{fig:SQUID}%
\end{figure}

%
%
%
Having demonstrated the high quality growth of TFO films in the orthorhombic phase on STO substrates, our attention now turns to the magnetic properties of the TFO thin films, which we investigated using superconducting quantum interference device (SQUID) magnetometry. FIG. \ref{fig:SQUID} (a) shows a field cooled curve and a field warming curve for which the sample has been measured in out-of-plane (oop) and in-plane (ip) configuration along the orthorhombic axes, respectively. Details of the method can be found in the appendix \ref{sec:MvsT}. The green region depicts the bulk spin reorientation region as measured in single crystals ($T_2=94$\,K and $T_1=82$\,K \cite{LEAKE1968}), where the N\'eel vector \textbf{N} smoothly rotates between the two magnetic states \cite{MagTFO2}. Above the SRT \textbf{N} is in-plane along the orthorhombic $a$-axis ($\Gamma_4$ phase). Since two crystallographic domains are present \textbf{N} should lie in both possible in-plane directions $45^\circ$ corresponding to the STO axes. In both cases \textbf{M} is pointing along the $c$-axis, which can be easily aligned in one direction using a magnetic field. Below the SRT the spins align along the $c$-axis and so \textbf{N} reorients out of the plane ($\Gamma_2$ phase). \textbf{M} can be directed in any of the 4 in-plane directions of the TFO $a$-axis. These two possible configurations are shown for their respective temperature ranges in FIG. \ref{fig:SQUID} (d). \\
Focusing first on the SQUID oop measurement, the measured magnetization at high temperatures corresponds to the canted moment of the TFO which increases slightly with decreasing temperature as can be explained by the parallel alignment of the paramagnetic Tm moments \cite{White1969} or an increased canting angle. While the first effect is observed in bulk, the latter effect is not \cite{MagTFO}, but could occur in thin films due to temperature dependent strain arising from different thermal expansion coefficients of TFO and STO. 
Across the SRT, both \textbf{N} and \textbf{M} rotate smoothly to a new equilibrium position. Indeed, we note that the oop magnetization reaches a maximum around $T_{II}=110\,$K, which we attribute to the start of the SRT. 
As we continue to reduce the temperature, we note that the oop magnetization continues to decrease as \textbf{M} no longer lies along this direction. We reach a minimum for the magnetization at around 50\,K, which stems from a superposition of the decreasing projection of \textbf{M} in $c$-direction and paramagnetic components increasing the signal at low temperatures. 
Therefore we make use of an ip measurement to determine the edge of the SRT. We observe a decreasing magnetic moment along the $a$-axis above $T_I=72$\,K when measuring along the substrate diagonal, which we attribute to the end of the SRT. \\These measurements indicate that the SRT is present in our thin films. This shows that they exhibit the desired magnetic properties as the SRT is similar to bulk samples. 
We note that $T_{I}=72\,$K and $T_{II}=110\,$K deviate slightly from the bulk values $T_1=82\,$K and $T_2=94\,$K. The extension of the SRT to $T_{1'}=82\,$K and $T_{2'}=120\,$K in near surface regions has been reported in Ref. \cite{Staub} for bulk samples. A further increase of the SRT region may stem from the small size of the TFO grains and the resulting enlarged surface area. There is also the probability that strain induced by the thin film growth alters the key magnetic anisotropies along the $a$ and $c$ axes responsible for the transition, as has been demonstrated for SmFeO$_3$ \cite{StrainSRT}. \par
%
%
%
%
To confirm our observations, we next perform measurements of the magnetization as a function of the applied magnetic field at a constant temperature above (200\,K) and below (40\,K) our observed SRT. Once the background from the diamagnetic substrate and paramagnetic components has been subtracted (see FIG. \ref{fig:SQUID} (b)), we observe a curve consisting of two contributions for an oop magnetic field at 200\,K: a broad hysteresis with a coercive field of 1.8\,T and a soft hysteresis at very low magnetic fields. For the 40\,K curve, the soft hysteresis around 0\,T is still present. A second contribution to the signal has adopted a shape more reminiscent of a hard axis ferromagnetic loop than an easy axis as expected. 
From low amplitude hysteresis loops discussed in the appendix section \ref{sec:minLoop} we observe that the soft-magnetic contribution is temperature independent and therefore not linked to the magnetic structure of the orthorhombic TFO.
Subtracting the soft magnetic contribution, the saturation magnetization at 200\,K is found to be around $0.06\,\mu$B/f.u., which is lower than what was recently reported on TFO bulk samples also using $M(H)$ measurements \cite{Zhou2020}. The magnitude of the saturation magnetization has two contributions: the paramagnetic Tm moments being polarized by the local crystal field generated by the Fe atoms and the canting of the Fe sublattices, which is related to the strength of the antisymmetric exchange interaction (Dzyaloshinskii Moriya interaction, DMI) present in all rare earth orthoferrites \cite{MagTFO}. The crystallographic distortion due to strain affects the crystal field, thus leading to deviations of the DMI through higher order terms of the spin-orbit coupling. Therefore, a deviation between magnetization values of bulk and thin films samples is not surprising. \\ 
%
%
%
We have demonstrated by SQUID measurements that our films have a measurable SRT signal. We have assumed that our samples fulfill the model of a canted antiferromagnet. However, we have yet to show that our films possess actually antiferromagnetic ordering to rule out the possibility that the SQUID signals stem from ferromagnetic impurities. In order to confirm that they are antiferromagnetic, x-ray absorption spectra were measured for circular and linear polarized photons around the Fe L$_2$ and L$_3$ edge at the Diamond Light Source I06 beamline. The magnetic circular dichroism (XMCD) and x-ray magnetic linear dichroism (XMLD) were calculated. Details on the method and the XMCD spectrum are given in the appendix section \ref{sec:XMLD}.
We note that there is no clear XMCD signal measured when using circularly polarized x-rays. The lack of XMCD means that in the near-surface region we do not observe evidence of a ferromagnetic phase or the canted moment. The missing of the latter is likely due the sensitivity limit of the technique. This is not surprising given its small magnitude and the finite sensitivity of the setup, as well as the difficulties in measuring out of plane magnetism with x-rays having a grazing incidence angle of $16^\circ$. In order to demonstrate antiferromagnetic nature of our films, we take x-ray absorption spectra (XAS) for horizontal and vertical linear polarized x-rays. The XMLD is calculated as XMLD=$I_{LH}-I_{LV}$, where $I_{LH}$ and $I_{LH}$ are the XAS of for horizontal and vertical linear polarization, respectively. The parallel (perpendicular) component of the N\'eel vector will give rise to a decrease (increase) of the XAS spectrum. We show in FIG. \ref{fig:SQUID} (c) the XAS and XMLD spectra calculated at 300\,K,  where clear dichroism is visible at both the Fe L$_2$ and L$_3$ edges.
The clear presence of XMLD and absence of XMCD demonstrates that our films are indeed antiferromagnetic.
The investigation of our TFO/STO thin films have shown that TFO grows in the orthorhombic phase like bulk crystals. These films show the typical TFO features like antiferromagnetic ordering and a SRT. However, we have so far used bulk techniques to detect this ordering. Device applications necessitate the all-electrical, on-chip detection of the AFM ordering and SRT, for which we turn to spin Hall magnetoresistance \cite{Nakayama2013}.\par
\begin{figure}[t]%
\centering
	\includegraphics[width=1\columnwidth]{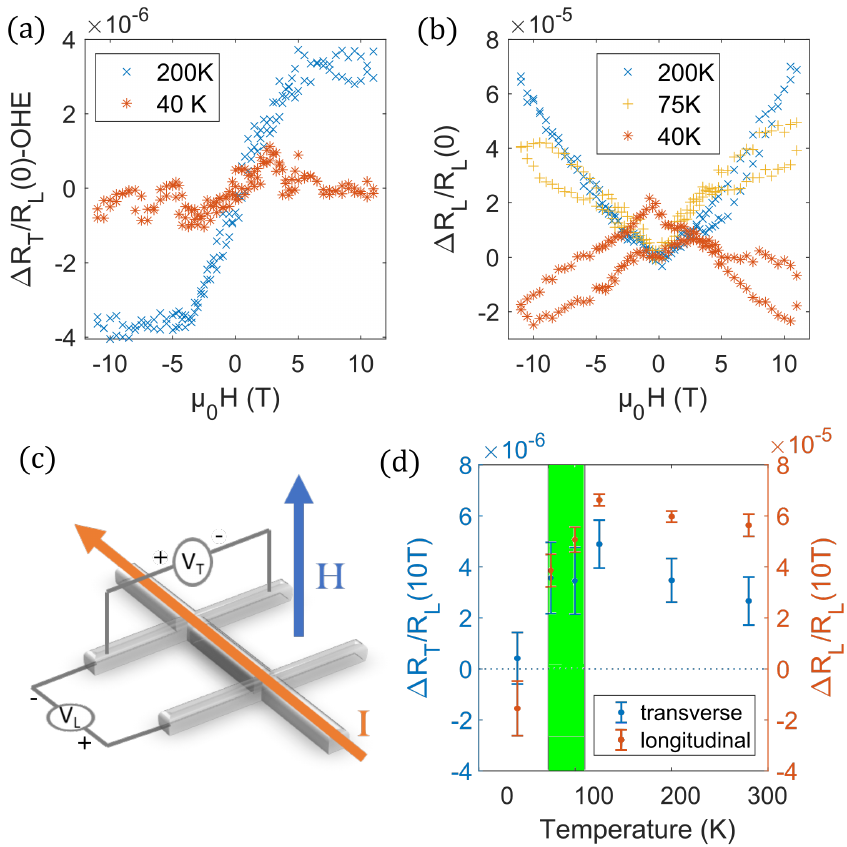}%
	\caption{(a) Transverse spin-Hall magnetoresistance (SMR) as a function of magnetic field at a temperature of 200\,K and 40\,K. The field is applied in $z$-direction (out of the plane). Note that the transverse SMR is extracted by subtracting the ordinary Hall effect from the measured transverse resistivity. A floating average has been applied with a range over 1\,T to smooth the data. (b) Relative longitudinal resistivity measured simultaneously. (c) Sketch of the measurement configuration. (d) Amplitude of transverse and longitudinal relative resistivity at 10\,T. In green the spin reorientation transition determined by SQUID measurements is depicted.}
	\label{fig:OOP}%
\end{figure}

%
%
%
%
%
Using lithographic methods, we define Pt Hall bars as detailed in the methods section \ref{sec:Methods} in order to electrically investigate the reorientation of the magnetic structure with temperature. The Hall bars are aligned parallel to the STO $a$-axis, and thus make a $45^\circ$ angle to the film $a$- and $b$-axes. $z$ is perpendicular to the Hall bar out of the plane, which coincides with the orthorhombic $c$-direction (see FIG. \ref{fig:OOP} (c)).\\ 
We pass a charge current along $x$ through the Pt, which leads to a spin accumulation $\mu_s$ at the Pt/TFO interface polarized along $y$, whilst capturing a longitudinal and transverse resistivity as a function of the magnetic field applied along $z$. The interaction between $\mu_s$ and the magnetic order parameters, which can be \textbf{N} \cite{Lebrun2019,NiOImaging} or \textbf{M} \cite{SMRSFO}, leads to a modulation of the Pt resistance due to the spin Hall magnetoresistance (SMR). Changes in the orientation of \textbf{N} and \textbf{M}, then lead to an electrically detectable response. The longitudinal resistivity can be plagued by numerous additional effects like ordinary magnetoresistance \cite{Isasa2016}, Hanle magnetoresistance \cite{Hanle} and sensitivity to small temperature changes, which may be comparable to, or even mask, the small effect of the SMR.
On the other hand, the interpretation of the transverse resistivity $R_T$ data is more straight forward to interpret. With a field applied in $z$ direction, $R_T$ has two key contributions: the ordinary Hall effect (OHE) \cite{Karplus1954} and the spin Hall anomalous Hall effect (SHAHE), which is related to the $z$-component of the magnetization of the material in contact with the conductor \cite{SMR}. 
Other effects like planar Hall effect \cite{Planar}, anomalous Hall effect \cite{Karplus1954}, magnetic SHE \cite{Kimata2019} and topological Hall effect \cite{Binz2008} are only relevant for magnetic conductors and can be discarded here where we consider an antiferromagnetic insulator. 
The OHE arises as a linearly increasing change of resistance with magnetic field making it simple to account for. In the case of an antiferromagnet without a significant canted moment, only the OHE is expected. However, we observe a distinct hysteresis in the transverse resistivity beneath the linear contribution from the OHE at 200\,K (FIG. \ref{fig:OOP} (a)). We note that the hysteresis saturates at a similar magnetic field as the hysteresis observed previously by SQUID. To determine further whether this hysteresis originates from the TFO films, we cool below the SRT to 40\,K, where the canted moment now lies in-plane. While the contribution from OHE remains unchanged, the hysteresis disappears, confirming that this signal originates from the net magnetic moment of the TFO films. This is shown in FIG. \ref{fig:OOP} (a). 
Given that, ignoring the N\'eel vector, the transverse resistivity is proportional to the $z$ component of the magnetization \cite{SMR,SMRYIG}, we attribute this hysteresis to the spin Hall anomalous Hall effect (SHAHE) \cite{SMR,SMR1} of the canted moment \textbf{M} oriented along $z$. 
Unlike in the SQUID measurements (FIG: \ref{fig:SQUID} (b)), a remnant magnetization in $R_T$ is not observed above the SRT. A fundamentally different spin configuration at the surface layer compared to bulk is not expected since previous experiments on bulk samples underline the similarity of surface sensitive and bulk measurements \cite{jetp1, Staub, SurfaceMagnetoelastic}. The absence of a remnant signal may arise from the broken inversion symmetry at the TFO/Pt interface competing with the bulk DMI, reducing the canted moment at the interface. The soft hysteresis, which may stem from a maghemite phase, has no obvious impact on the measurement.
We also note that the signal has the same symmetry as proximity induced AHE. However, we do not expect proximity effect in our samples as also not observed in other Pt/AFM heterostructures \cite{NioProx} given the lack of a large net magnetic moment and stray fields. 
We therefore suggest that the transverse signal is dominated by the orientation of \textbf{M}, specifically the $z$ component, reminiscent of the SHAHE observed in ferromagnets. The dominance of the magnetic spin mixing conductance in the transverse signal would indicate that the magnetic spin mixing conductance should also dominate the longitudinal response FIG \ref{fig:OOP} (b).
In fact, we do see a significant change in the longitudinal resistivity as a function of magnetic field when going though the SRT. The amplitude clearly decreases with decreasing temperature around the SRT. Within the transition we see an intermediate state which might indicate the magnetic field induced reorientation of \textbf{M} and \textbf{N} \cite{LeCraw1968}. This would indicate a $R_L$-dependence on \textbf{M}, having a positive magnetoresistive (MR) effect together with a decrease of $M_y$ and an increase of the $N_y$ contribution. Assuming this dependence and ignoring other effects, one would expect no MR effect at 200\,K, because \textbf{M} does not have a $y$-contribution above the SRT. However, we  observe a large positive MR that could be explained by the dependence of $R_L$ in \textbf{N}. Therefore, clearly identifying the dependence of $R_L$ on \textbf{M} or \textbf{N} is challenging and need more measurements, also to identify the contributions from temperature and other MR effects, i.e. OMR and Hanle MR, which have the same symmetry as SMR in our measurement geometry. \\
In summary, we observe a strong change of both $R_T$ and $R_L$ through the SRT. The dependence of the signal amplitude with temperature is shown in FIG.\ref{fig:OOP} (d). 
The amplitude of $R_T$ follows our expectations centered around a dominant contribution from $M_z$ and the associated SHAHE, however, as discussed previously, the dependence of $R_L$ cannot be explained through the $y$ component of either \textbf{N} or \textbf{M}. Instead, the behaviour appears to be in a certain sense related to the out of plane component of the magnetization. However, this is not consistent with the description of SMR \cite{Nakayama2013,SMR} and would require further study of the SMR in canted AFMs with a significant DMI that goes beyond the scope of this work. 

\section{Conclusion} 
We have successfully grown samples of oriented orthorhombic TmFeO$_3$ thin films on SrTiO$_3$ substrates by pulsed laser deposition. We observe a four-fold symmetry of the in-plane ordering due to the cubic symmetry of the substrate resulting in crystallographic twinning. The samples exhibit antiferromagnetic ordering, a canted moment and a spin reorientation transition (SRT), as confirmed by SQUID magnetometry and XMLD. The SRT is shifted compared to bulk samples due to strained growth altering the local crystal field. To detect the SRT electrically, we utilize spin Hall magnetoresistance (SMR). We can ascertain the different magnetic phases utilizing SHAHE and also observe anomalies in the longitudinal SMR signal going through the SRT. The surface sensitive nature of SMR overcomes the need for bulk measurements that can, for thinner films, be difficult due to the small volume and substrate contributions. Our results demonstrate, that one can electrically detect the spin reorientation transition of canted antiferromagnetic orthoferrites, which is important for future spintronic applications such as memory devices and canted antiferromagnetic nanooscillators.

\section*{Acknowledgments}
The authors gratefully acknowledge funding by the Deutsche
Forschungsgemeinschaft (DFG, German Research Foundation) – project
number 35867137. This work was supported by the Max Planck Graduate Center with the Johannes Gutenberg-Universität Mainz (MPGC) as well the Graduate School of Excellence Material Science in Mainz (GSC266). Funded by the Deutsche Forschungsgemeinschaft (DFG, German Research Foundation) – TRR
173–268565370 and KAUST COSR-2019-CR68-4048.2. L.B acknowledges the European Union’s Horizon 2020 research and innovation program under the Marie Sk\l{}odowska-Curie grant agreement ARTES  number 793159. We acknowledge Diamond Light Source for time on beam line I06 under proposal MM23819-1. 

\section{Appendix}
\subsection{Methods}
\label{sec:Methods}
A TmFeO$_3$ target has been prepared by standard solid state reaction out of Tm$_2$O$_3$ and Fe$_2$O$_3$ powder.
Thin films have been prepared by pulsed laser deposition (PLD) on SrTiO$_3$ (001) substrates using a KrF Compex Pro 205 excimer laser and a vacuum chamber with a base pressure of $2\cdot10^{-8}$\,mbar. 
\begin{table}[ht]
	\centering
		\begin{tabular}{|c|c|}
			\hline
			Parameter														& value \\\hline
			Deposition temperature ($^\circ$C)	&	630 \\
			O$_2$ pressure (mbar)								&	0.2 \\
			Substrate-target distance (cm)			&	5.5 \\
			Laser energy per pulse (mJ)					& 130 \\
			Laser spot size (mm$^2$)						&	9 \\
			Laser pulse frequency (Hz)					&	10 \\
			Cooling rate after dep. (K/min)			&	25	\\
			Growth rate (nm/min)								&	1.6	\\\hline
		\end{tabular}
	\caption{Deposition conditions for PLD of TFO thin films}
	\label{tab:PLD}
\end{table}
The structural properties of the samples have been investigated using x-ray diffraction (XRD) and a Bruker D8 four circle diffractometer. The temperature dependent magnetic properties have been analyzed using a superconducting quantum interference device (SQUID) from Quantum Design. The surface morphology has been imaged using atomic force microscopy (AFM) with a Digital Instruments 3100 Dimension equipped with a NanoScope IV controller. As probes, Bruker SNL-10 have been used. For plotting the images and calculating the grain size Gwyddion software \cite{Gwyddion} was utilized.\\
For the transport measurements, platinum Hall bars have been defined by lithographic methods, with Pt deposited using DC magnetron sputtering at room temperature and a lift-off process. The dimensions of the Hall bars are $10\,\mu$m in width, 7\,nm in thickness and $100\,\mu$m in length. The arms are separated by $55\,\mu$m and have a width of $3\,\mu$m. Electrical measurements at a Hall bar have been performed in a cryostat equipped with a variable temperature insert. Magnetic fields up to 11\,T and temperatures in the range of 5\,K to 300\,K can be accessed. An identical sample as utilized for transport measurements was covered by 2\,nm of Pt in order to prevent charging and measured at DIAMOND beam line I06. Before measuring the x-ray absorption spectra (XAS) the sample has been put in a magnetic field of 5\,T applied in $c$-direction. XAS have been measured around the L$_3$ and L$_2$ iron absorption edge.

\subsection{AFM}
\label{sec:AFM}%

\begin{figure}[b]%
\centering
	\includegraphics[width=.5\columnwidth]{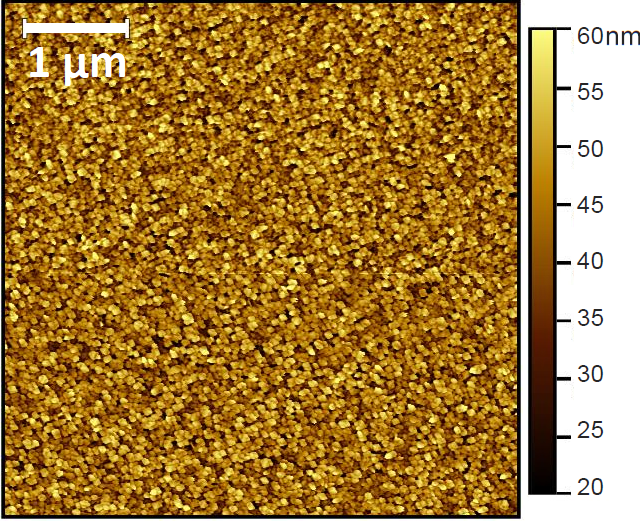}%
	\caption{Atomic force microscopy image of the sample surface. The scale bar corresponds to 1\,$\mu$m.}
	\label{fig:AFM}%
\end{figure}
The surface of the sample has been investigated by atomic force microscopy. The data of a 5-by-5\,$\mu$m scan has been analyzed with Gwyddion software \cite{Gwyddion}. An image of the topography is shown in FIG. \ref{fig:AFM}. We determine an overall root mean square (RMS) roughness ($Sq$) of around 8\,nm for a film thickness of 200\,nm. The size of the surface grains has been calculated by masking the grains using a height threshold of 43\%, resulting in a mean grain size of 42\,nm. 
We suggest that the grain size could be increased by making use of substrates without the cubic symmetry of STO to prevent the formation of twinned domains (e.g. DyScO$_3$, NdGaO$_3$, YAlO$_3$ or related orthorhombic materials). Whilst this is not of relevance for the work here, where the transport measurements are integrating over areas far larger than the individual grain size, other branches of antiferromagnetic spintronics can benefit from a reduction in grain and magnetic domain boundaries \cite{transport}.\par

\subsection{SQUID M vs T curves}
\label{sec:MvsT}
The measurement of the SRT has been performed in the SQUID, measuring the temperature dependent magnetization along the TFO $c$- and $a$-axis as shown in the main text. 
For the oop measurement (along the $c$-axis) the sample was heated up to 390\,K. At this temperature a field magnitude of 5\,T was briefly applied along the $c$-axis to align \textbf{M} of the domains. Then a field of 30\,mT was applied and the temperature dependent magnetization was measured during cool-down at a rate of 2\,K/min. 
For the ip measurement, a magnetic field of 5\,T was applied along the $a/b$ TFO axis, $45^\circ$ to the STO axes, at a temperature of 20\,K for a short period of time, again to saturate the canted moment. The sample was then heated in a magnetic field of 50\,mT at 2\,K/min. Here, a larger magnetic field has been chosen because at lower fields ferro-, para- and diamagnetic contributions compensate at some point in the measurement range. This leads to the measurement software not being able to fit the raw signal which further leads to jumps in the $M(T)$ curve that are not originating from the sample's properties.

\subsection{SQUID minor loops} 
\label{sec:minLoop}
\begin{figure}[t]%
\centering
	\includegraphics[width=.5\columnwidth]{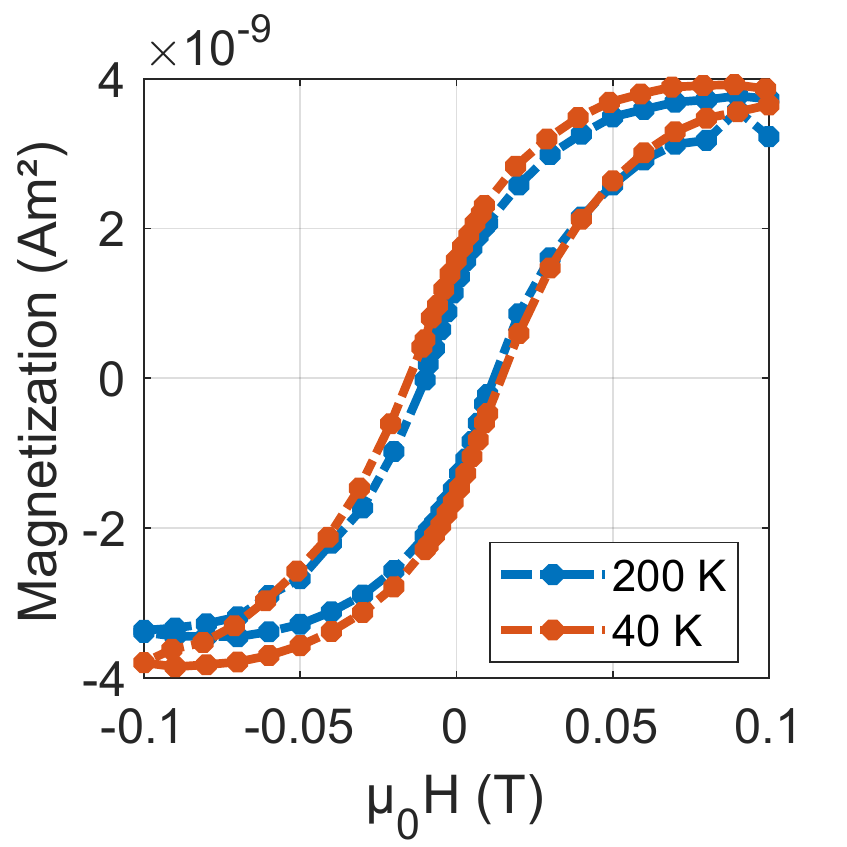}%
	\caption{Minor loop measured in SQUID in out-of-plane configuration at different temperatures with a maximum applied field of 0.1\,T. Note that the scale of can not be normalized because of the unknown volume of the soft magnetic phase. The data has been corrected for para- and diamagnetic background.}
	\label{fig:Minor}%
\end{figure}

We investigate the soft magnetic phase by taking minor loops around 0\,T so as to exclude the high field contributions. This is done by first applying a field of -5\,T to the sample for a short period of time. Then the $M(H)$ loop is measured with $H_{max}=0.1\,$T starting from positive values. 
We observe a hysteresis that has a coercive field of 12\,mT (18\,mT) and a maximum magnetic moment of $3.6\cdot10^{-9}$\,Am$^2$ ($3.8\cdot10^{-9}$\,Am$^2$) at a temperature of 200\,K (40\,K).
This contribution does not change significantly with temperature as shown in FIG. \ref{fig:Minor}. Note that the values for the magnetization are not normalized because the volume of the soft magnetic phase is unknown.
Hysteresis curves with a soft and hard magnetic contribution are in literature referred to as wasp-waisted hysteresis loops. They have already been reported for other oxides like CoFe$_2$O$_4$ thin films \cite{CFO1,CFO2,CFO3} and Fe$_3$O$_4$/CoO bilayers \cite{CFO4}. For polycristalline orthoferrite thin films (not TFO), there are reports of strange phases leading to a soft hysteresis \cite{Garnet,Magnetite}. Possible phases which are named are a garnet phase, magnetite (Fe$_3$O$_4$) and rare earth oxide. For TFO the corresponding garnet phase is Tm$_3$Fe$_5$O$_{12}$ (TIG), an insulating ferrimagnet, whose magnetization changes by a factor of around 3 between 200\,K and 40\,K \cite{Pauthenet1958}. Magnetite possesses a phase transition at 125\,K below witch the coercivity increases drastically \cite{zdemir2002} and Tm$_2$O$_3$ is a paramagnetic material \cite{Gondek2010}. So none of these candidates seem to explain the similarity of the soft magnetic contributions at 40\,K and 200\,K in our TFO thin films. For hematite ($\alpha$-Fe$_2$O$_3$), a related component to orthoferrites, thin films possess a thin maghemite ($\gamma$-Fe$_2$O$_2$) layer at the substrate interface \cite{Barbier2005}. If we assume such a layer in our sample the magnetic signal corresponds to a maghemite volume fraction of 0.2\% of the whole TFO layer thickness. 
We cannot exclude the presence of this phase from XRD measurements since the count rate of such a small volume is around the noise level in our setup, even if it was well oriented. 
Nevertheless, we see that the impact on electric measurements by the soft magnetic phase, whether maghemite or not, is negligible for our studies. \\

\subsection{XMLD}
\label{sec:XMLD}
\begin{figure}[b]%
\centering
	\includegraphics[width=0.5\columnwidth]{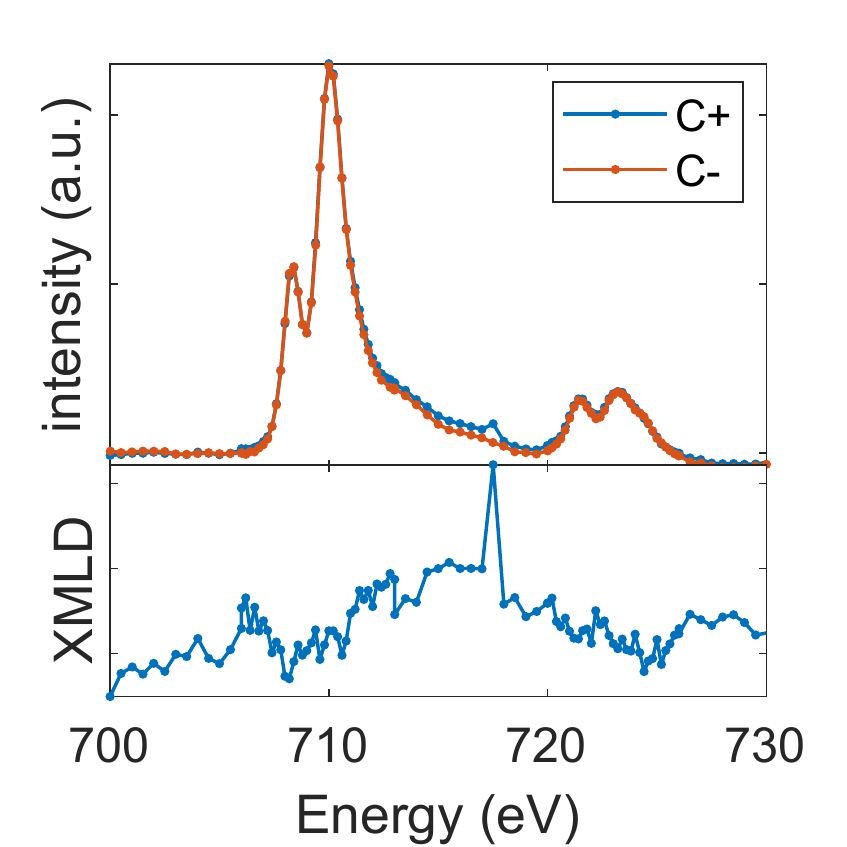}%
	\caption{X-ray absorption spectra for circular polarized photons C+ and C- (top). XMCD spectrum calculated as the difference between C+ and C- (bottom).}
	\label{fig:XMCD}
\end{figure}

In order to confirm that our TFO thin films are antiferromagnetic, x-ray absorption spectra (XAS) where captured at the Fe $L_{2/3}$ edge for linear horizontal (LH) and vertical (LV) polarization as well as positive (C+) and negative circular (C-) polarization. The sample has been aligned with the orthorhombic axes in the reflection plane. The incident angle of the beam is $16^\circ$. The data was processed as following: The reflected intensity was divided by the incident intensity of the x-ray beam to account for modulation of the incoming intensity. The background away from the $L_2$ and $L_3$ peak was subtracted to shift the baseline to zero of each curve. A correction factor has been multiplied to LH and C+ to align LH and LV as well as C+ and C- curves. XMLD and XMCD are calculated by XMLD=$I_{LH}-I_{LV}$ and XMCD=$I_{C+}-I_{C-}$. The XMCD spectrum is shown in FIG. \ref{fig:XMCD}, while the XMLD spectrum is shown in the main text. %
While for the XMLD spectrum we observe a clear signal around the $L_3$ edge, there is no clear signal to be identified in the XMCD spectrum. The lack of XMCD means that in the near-surface region we do not observe evidence of a ferromagnetic phase or the canted moment.

\bibliography{BibTex1}

\begin{thebibliography}{52}%
\makeatletter
\providecommand \@ifxundefined [1]{%
 \@ifx{#1\undefined}
}%
\providecommand \@ifnum [1]{%
 \ifnum #1\expandafter \@firstoftwo
 \else \expandafter \@secondoftwo
 \fi
}%
\providecommand \@ifx [1]{%
 \ifx #1\expandafter \@firstoftwo
 \else \expandafter \@secondoftwo
 \fi
}%
\providecommand \natexlab [1]{#1}%
\providecommand \enquote  [1]{``#1''}%
\providecommand \bibnamefont  [1]{#1}%
\providecommand \bibfnamefont [1]{#1}%
\providecommand \citenamefont [1]{#1}%
\providecommand \href@noop [0]{\@secondoftwo}%
\providecommand \href [0]{\begingroup \@sanitize@url \@href}%
\providecommand \@href[1]{\@@startlink{#1}\@@href}%
\providecommand \@@href[1]{\endgroup#1\@@endlink}%
\providecommand \@sanitize@url [0]{\catcode `\\12\catcode `\$12\catcode
  `\&12\catcode `\#12\catcode `\^12\catcode `\_12\catcode `\%12\relax}%
\providecommand \@@startlink[1]{}%
\providecommand \@@endlink[0]{}%
\providecommand \url  [0]{\begingroup\@sanitize@url \@url }%
\providecommand \@url [1]{\endgroup\@href {#1}{\urlprefix }}%
\providecommand \urlprefix  [0]{URL }%
\providecommand \Eprint [0]{\href }%
\providecommand \doibase [0]{https://doi.org/}%
\providecommand \selectlanguage [0]{\@gobble}%
\providecommand \bibinfo  [0]{\@secondoftwo}%
\providecommand \bibfield  [0]{\@secondoftwo}%
\providecommand \translation [1]{[#1]}%
\providecommand \BibitemOpen [0]{}%
\providecommand \bibitemStop [0]{}%
\providecommand \bibitemNoStop [0]{.\EOS\space}%
\providecommand \EOS [0]{\spacefactor3000\relax}%
\providecommand \BibitemShut  [1]{\csname bibitem#1\endcsname}%
\let\auto@bib@innerbib\@empty
\bibitem [{\citenamefont {Baltz}\ \emph {et~al.}(2018)\citenamefont {Baltz},
  \citenamefont {Manchon}, \citenamefont {Tsoi}, \citenamefont {Moriyama},
  \citenamefont {Ono},\ and\ \citenamefont
  {Tserkovnyak}}]{RevModPhys.90.015005}%
  \BibitemOpen
  \bibfield  {author} {\bibinfo {author} {\bibfnamefont {V.}~\bibnamefont
  {Baltz}}, \bibinfo {author} {\bibfnamefont {A.}~\bibnamefont {Manchon}},
  \bibinfo {author} {\bibfnamefont {M.}~\bibnamefont {Tsoi}}, \bibinfo {author}
  {\bibfnamefont {T.}~\bibnamefont {Moriyama}}, \bibinfo {author}
  {\bibfnamefont {T.}~\bibnamefont {Ono}},\ and\ \bibinfo {author}
  {\bibfnamefont {Y.}~\bibnamefont {Tserkovnyak}},\ }\bibfield  {title}
  {\bibinfo {title} {Antiferromagnetic spintronics},\ }\href
  {https://doi.org/10.1103/RevModPhys.90.015005} {\bibfield  {journal}
  {\bibinfo  {journal} {Rev. Mod. Phys.}\ }\textbf {\bibinfo {volume} {90}},\
  \bibinfo {pages} {015005} (\bibinfo {year} {2018})}\BibitemShut {NoStop}%
\bibitem [{\citenamefont {Jungwirth}\ \emph {et~al.}(2016)\citenamefont
  {Jungwirth}, \citenamefont {Marti}, \citenamefont {Wadley},\ and\
  \citenamefont {Wunderlich}}]{Jungwirth2016}%
  \BibitemOpen
  \bibfield  {author} {\bibinfo {author} {\bibfnamefont {T.}~\bibnamefont
  {Jungwirth}}, \bibinfo {author} {\bibfnamefont {X.}~\bibnamefont {Marti}},
  \bibinfo {author} {\bibfnamefont {P.}~\bibnamefont {Wadley}},\ and\ \bibinfo
  {author} {\bibfnamefont {J.}~\bibnamefont {Wunderlich}},\ }\bibfield  {title}
  {\bibinfo {title} {Antiferromagnetic spintronics},\ }\href
  {https://doi.org/10.1038/nnano.2016.18} {\bibfield  {journal} {\bibinfo
  {journal} {Nature Nanotechnology}\ }\textbf {\bibinfo {volume} {11}},\
  \bibinfo {pages} {231} (\bibinfo {year} {2016})}\BibitemShut {NoStop}%
\bibitem [{\citenamefont {Jungwirth}\ \emph {et~al.}(2018)\citenamefont
  {Jungwirth}, \citenamefont {Sinova}, \citenamefont {Manchon}, \citenamefont
  {Marti}, \citenamefont {Wunderlich},\ and\ \citenamefont
  {Felser}}]{Jungwirth2018}%
  \BibitemOpen
  \bibfield  {author} {\bibinfo {author} {\bibfnamefont {T.}~\bibnamefont
  {Jungwirth}}, \bibinfo {author} {\bibfnamefont {J.}~\bibnamefont {Sinova}},
  \bibinfo {author} {\bibfnamefont {A.}~\bibnamefont {Manchon}}, \bibinfo
  {author} {\bibfnamefont {X.}~\bibnamefont {Marti}}, \bibinfo {author}
  {\bibfnamefont {J.}~\bibnamefont {Wunderlich}},\ and\ \bibinfo {author}
  {\bibfnamefont {C.}~\bibnamefont {Felser}},\ }\bibfield  {title} {\bibinfo
  {title} {The multiple directions of antiferromagnetic spintronics},\ }\href
  {https://doi.org/10.1038/s41567-018-0063-6} {\bibfield  {journal} {\bibinfo
  {journal} {Nature Physics}\ }\textbf {\bibinfo {volume} {14}},\ \bibinfo
  {pages} {200} (\bibinfo {year} {2018})}\BibitemShut {NoStop}%
\bibitem [{\citenamefont {Sulymenko}\ \emph {et~al.}(2017)\citenamefont
  {Sulymenko}, \citenamefont {Prokopenko}, \citenamefont {Tiberkevich},
  \citenamefont {Slavin}, \citenamefont {Ivanov},\ and\ \citenamefont
  {Khymyn}}]{Sulymenko2017}%
  \BibitemOpen
  \bibfield  {author} {\bibinfo {author} {\bibfnamefont {O.~R.}\ \bibnamefont
  {Sulymenko}}, \bibinfo {author} {\bibfnamefont {O.~V.}\ \bibnamefont
  {Prokopenko}}, \bibinfo {author} {\bibfnamefont {V.~S.}\ \bibnamefont
  {Tiberkevich}}, \bibinfo {author} {\bibfnamefont {A.~N.}\ \bibnamefont
  {Slavin}}, \bibinfo {author} {\bibfnamefont {B.~A.}\ \bibnamefont {Ivanov}},\
  and\ \bibinfo {author} {\bibfnamefont {R.~S.}\ \bibnamefont {Khymyn}},\
  }\bibfield  {title} {\bibinfo {title} {Terahertz-frequency spin hall
  auto-oscillator based on a canted antiferromagnet},\ }\href
  {https://doi.org/10.1103/PhysRevApplied.8.064007} {\bibfield  {journal}
  {\bibinfo  {journal} {Phys. Rev. Applied}\ }\textbf {\bibinfo {volume} {8}},\
  \bibinfo {pages} {064007} (\bibinfo {year} {2017})}\BibitemShut {NoStop}%
\bibitem [{\citenamefont {Stremoukhov}\ \emph {et~al.}(2019)\citenamefont
  {Stremoukhov}, \citenamefont {Safin}, \citenamefont {Logunov}, \citenamefont
  {Nikitov},\ and\ \citenamefont {Kirilyuk}}]{Stremoukhov2019}%
  \BibitemOpen
  \bibfield  {author} {\bibinfo {author} {\bibfnamefont {P.}~\bibnamefont
  {Stremoukhov}}, \bibinfo {author} {\bibfnamefont {A.}~\bibnamefont {Safin}},
  \bibinfo {author} {\bibfnamefont {M.}~\bibnamefont {Logunov}}, \bibinfo
  {author} {\bibfnamefont {S.}~\bibnamefont {Nikitov}},\ and\ \bibinfo {author}
  {\bibfnamefont {A.}~\bibnamefont {Kirilyuk}},\ }\bibfield  {title} {\bibinfo
  {title} {Spintronic terahertz-frequency nonlinear emitter based on the canted
  antiferromagnet-platinum bilayers},\ }\href
  {https://doi.org/10.1063/1.5090455} {\bibfield  {journal} {\bibinfo
  {journal} {Journal of Applied Physics}\ }\textbf {\bibinfo {volume} {125}},\
  \bibinfo {pages} {223903} (\bibinfo {year} {2019})}\BibitemShut {NoStop}%
\bibitem [{\citenamefont {Wadley}\ \emph {et~al.}(2016)\citenamefont {Wadley},
  \citenamefont {Howells}, \citenamefont {elezny}, \citenamefont {Andrews},
  \citenamefont {Hills}, \citenamefont {Campion}, \citenamefont {Novak},
  \citenamefont {Olejnik}, \citenamefont {Maccherozzi}, \citenamefont {Dhesi},
  \citenamefont {Martin}, \citenamefont {Wagner}, \citenamefont {Wunderlich},
  \citenamefont {Freimuth}, \citenamefont {Mokrousov}, \citenamefont {Kune},
  \citenamefont {Chauhan}, \citenamefont {Grzybowski}, \citenamefont
  {Rushforth}, \citenamefont {Edmonds}, \citenamefont {Gallagher},\ and\
  \citenamefont {Jungwirth}}]{Wadley2016}%
  \BibitemOpen
  \bibfield  {author} {\bibinfo {author} {\bibfnamefont {P.}~\bibnamefont
  {Wadley}}, \bibinfo {author} {\bibfnamefont {B.}~\bibnamefont {Howells}},
  \bibinfo {author} {\bibfnamefont {J.}~\bibnamefont {elezny}}, \bibinfo
  {author} {\bibfnamefont {C.}~\bibnamefont {Andrews}}, \bibinfo {author}
  {\bibfnamefont {V.}~\bibnamefont {Hills}}, \bibinfo {author} {\bibfnamefont
  {R.~P.}\ \bibnamefont {Campion}}, \bibinfo {author} {\bibfnamefont
  {V.}~\bibnamefont {Novak}}, \bibinfo {author} {\bibfnamefont
  {K.}~\bibnamefont {Olejnik}}, \bibinfo {author} {\bibfnamefont
  {F.}~\bibnamefont {Maccherozzi}}, \bibinfo {author} {\bibfnamefont {S.~S.}\
  \bibnamefont {Dhesi}}, \bibinfo {author} {\bibfnamefont {S.~Y.}\ \bibnamefont
  {Martin}}, \bibinfo {author} {\bibfnamefont {T.}~\bibnamefont {Wagner}},
  \bibinfo {author} {\bibfnamefont {J.}~\bibnamefont {Wunderlich}}, \bibinfo
  {author} {\bibfnamefont {F.}~\bibnamefont {Freimuth}}, \bibinfo {author}
  {\bibfnamefont {Y.}~\bibnamefont {Mokrousov}}, \bibinfo {author}
  {\bibfnamefont {J.}~\bibnamefont {Kune}}, \bibinfo {author} {\bibfnamefont
  {J.~S.}\ \bibnamefont {Chauhan}}, \bibinfo {author} {\bibfnamefont {M.~J.}\
  \bibnamefont {Grzybowski}}, \bibinfo {author} {\bibfnamefont {A.~W.}\
  \bibnamefont {Rushforth}}, \bibinfo {author} {\bibfnamefont {K.~W.}\
  \bibnamefont {Edmonds}}, \bibinfo {author} {\bibfnamefont {B.~L.}\
  \bibnamefont {Gallagher}},\ and\ \bibinfo {author} {\bibfnamefont
  {T.}~\bibnamefont {Jungwirth}},\ }\bibfield  {title} {\bibinfo {title}
  {Electrical switching of an antiferromagnet},\ }\href
  {https://doi.org/10.1126/science.aab1031} {\bibfield  {journal} {\bibinfo
  {journal} {Science}\ }\textbf {\bibinfo {volume} {351}},\ \bibinfo {pages}
  {587} (\bibinfo {year} {2016})}\BibitemShut {NoStop}%
\bibitem [{\citenamefont {Schlauderer}\ \emph {et~al.}(2019)\citenamefont
  {Schlauderer}, \citenamefont {Lange}, \citenamefont {Baierl}, \citenamefont
  {Ebnet}, \citenamefont {Schmid}, \citenamefont {Valovcin}, \citenamefont
  {Zvezdin}, \citenamefont {Kimel}, \citenamefont {Mikhaylovskiy},\ and\
  \citenamefont {Huber}}]{Schlauderer2019}%
  \BibitemOpen
  \bibfield  {author} {\bibinfo {author} {\bibfnamefont {S.}~\bibnamefont
  {Schlauderer}}, \bibinfo {author} {\bibfnamefont {C.}~\bibnamefont {Lange}},
  \bibinfo {author} {\bibfnamefont {S.}~\bibnamefont {Baierl}}, \bibinfo
  {author} {\bibfnamefont {T.}~\bibnamefont {Ebnet}}, \bibinfo {author}
  {\bibfnamefont {C.~P.}\ \bibnamefont {Schmid}}, \bibinfo {author}
  {\bibfnamefont {D.~C.}\ \bibnamefont {Valovcin}}, \bibinfo {author}
  {\bibfnamefont {A.~K.}\ \bibnamefont {Zvezdin}}, \bibinfo {author}
  {\bibfnamefont {A.~V.}\ \bibnamefont {Kimel}}, \bibinfo {author}
  {\bibfnamefont {R.~V.}\ \bibnamefont {Mikhaylovskiy}},\ and\ \bibinfo
  {author} {\bibfnamefont {R.}~\bibnamefont {Huber}},\ }\bibfield  {title}
  {\bibinfo {title} {Temporal and spectral fingerprints of ultrafast
  all-coherent spin switching},\ }\href
  {https://doi.org/10.1038/s41586-019-1174-7} {\bibfield  {journal} {\bibinfo
  {journal} {Nature}\ }\textbf {\bibinfo {volume} {569}},\ \bibinfo {pages}
  {383} (\bibinfo {year} {2019})}\BibitemShut {NoStop}%
\bibitem [{\citenamefont {Jourdan}\ \emph {et~al.}(2015)\citenamefont
  {Jourdan}, \citenamefont {Br\"{a}uning}, \citenamefont {Sapozhnik},
  \citenamefont {Elmers}, \citenamefont {Zabel},\ and\ \citenamefont
  {Kl\"{a}ui}}]{Jourdan2015}%
  \BibitemOpen
  \bibfield  {author} {\bibinfo {author} {\bibfnamefont {M.}~\bibnamefont
  {Jourdan}}, \bibinfo {author} {\bibfnamefont {H.}~\bibnamefont
  {Br\"{a}uning}}, \bibinfo {author} {\bibfnamefont {A.}~\bibnamefont
  {Sapozhnik}}, \bibinfo {author} {\bibfnamefont {H.-J.}\ \bibnamefont
  {Elmers}}, \bibinfo {author} {\bibfnamefont {H.}~\bibnamefont {Zabel}},\ and\
  \bibinfo {author} {\bibfnamefont {M.}~\bibnamefont {Kl\"{a}ui}},\ }\bibfield
  {title} {\bibinfo {title} {Epitaxial {Mn}$_2${Au} thin films for
  antiferromagnetic spintronics},\ }\href
  {https://doi.org/10.1088/0022-3727/48/38/385001} {\bibfield  {journal}
  {\bibinfo  {journal} {Journal of Physics D: Applied Physics}\ }\textbf
  {\bibinfo {volume} {48}},\ \bibinfo {pages} {385001} (\bibinfo {year}
  {2015})}\BibitemShut {NoStop}%
\bibitem [{\citenamefont {Mix}\ and\ \citenamefont {Jakob}(2013)}]{Mix2013}%
  \BibitemOpen
  \bibfield  {author} {\bibinfo {author} {\bibfnamefont {C.}~\bibnamefont
  {Mix}}\ and\ \bibinfo {author} {\bibfnamefont {G.}~\bibnamefont {Jakob}},\
  }\bibfield  {title} {\bibinfo {title} {Multiferroic and structural properties
  of {BiFeO}$_3$ close to the strain induced phase transition on different
  substrates},\ }\href {https://doi.org/10.1063/1.4795216} {\bibfield
  {journal} {\bibinfo  {journal} {Journal of Applied Physics}\ }\textbf
  {\bibinfo {volume} {113}},\ \bibinfo {pages} {17D907} (\bibinfo {year}
  {2013})}\BibitemShut {NoStop}%
\bibitem [{\citenamefont {Baldrati}\ \emph {et~al.}(2019)\citenamefont
  {Baldrati}, \citenamefont {Gomonay}, \citenamefont {Ross}, \citenamefont
  {Filianina}, \citenamefont {Lebrun}, \citenamefont {Ramos}, \citenamefont
  {Leveille}, \citenamefont {Fuhrmann}, \citenamefont {Forrest}, \citenamefont
  {Maccherozzi}, \citenamefont {Valencia}, \citenamefont {Kronast},
  \citenamefont {Saitoh}, \citenamefont {Sinova},\ and\ \citenamefont
  {Kl\"aui}}]{NiOImaging}%
  \BibitemOpen
  \bibfield  {author} {\bibinfo {author} {\bibfnamefont {L.}~\bibnamefont
  {Baldrati}}, \bibinfo {author} {\bibfnamefont {O.}~\bibnamefont {Gomonay}},
  \bibinfo {author} {\bibfnamefont {A.}~\bibnamefont {Ross}}, \bibinfo {author}
  {\bibfnamefont {M.}~\bibnamefont {Filianina}}, \bibinfo {author}
  {\bibfnamefont {R.}~\bibnamefont {Lebrun}}, \bibinfo {author} {\bibfnamefont
  {R.}~\bibnamefont {Ramos}}, \bibinfo {author} {\bibfnamefont
  {C.}~\bibnamefont {Leveille}}, \bibinfo {author} {\bibfnamefont
  {F.}~\bibnamefont {Fuhrmann}}, \bibinfo {author} {\bibfnamefont {T.~R.}\
  \bibnamefont {Forrest}}, \bibinfo {author} {\bibfnamefont {F.}~\bibnamefont
  {Maccherozzi}}, \bibinfo {author} {\bibfnamefont {S.}~\bibnamefont
  {Valencia}}, \bibinfo {author} {\bibfnamefont {F.}~\bibnamefont {Kronast}},
  \bibinfo {author} {\bibfnamefont {E.}~\bibnamefont {Saitoh}}, \bibinfo
  {author} {\bibfnamefont {J.}~\bibnamefont {Sinova}},\ and\ \bibinfo {author}
  {\bibfnamefont {M.}~\bibnamefont {Kl\"aui}},\ }\bibfield  {title} {\bibinfo
  {title} {Mechanism of {N\'eel} order switching in antiferromagnetic thin
  films revealed by magnetotransport and direct imaging},\ }\href
  {https://doi.org/10.1103/PhysRevLett.123.177201} {\bibfield  {journal}
  {\bibinfo  {journal} {Phys. Rev. Lett.}\ }\textbf {\bibinfo {volume} {123}},\
  \bibinfo {pages} {177201} (\bibinfo {year} {2019})}\BibitemShut {NoStop}%
\bibitem [{\citenamefont {Moriyama}\ \emph {et~al.}(2018)\citenamefont
  {Moriyama}, \citenamefont {Oda}, \citenamefont {Ohkochi}, \citenamefont
  {Kimata},\ and\ \citenamefont {Ono}}]{Moriyama2018}%
  \BibitemOpen
  \bibfield  {author} {\bibinfo {author} {\bibfnamefont {T.}~\bibnamefont
  {Moriyama}}, \bibinfo {author} {\bibfnamefont {K.}~\bibnamefont {Oda}},
  \bibinfo {author} {\bibfnamefont {T.}~\bibnamefont {Ohkochi}}, \bibinfo
  {author} {\bibfnamefont {M.}~\bibnamefont {Kimata}},\ and\ \bibinfo {author}
  {\bibfnamefont {T.}~\bibnamefont {Ono}},\ }\bibfield  {title} {\bibinfo
  {title} {Spin torque control of antiferromagnetic moments in {NiO}},\ }\href
  {https://doi.org/10.1038/s41598-018-32508-w} {\bibfield  {journal} {\bibinfo
  {journal} {Scientific Reports}\ }\textbf {\bibinfo {volume} {8}},\ \bibinfo
  {pages} {14167} (\bibinfo {year} {2018})}\BibitemShut {NoStop}%
\bibitem [{\citenamefont {Ahn}\ \emph {et~al.}(2014)\citenamefont {Ahn},
  \citenamefont {Lee}, \citenamefont {Jang},\ and\ \citenamefont
  {Jeong}}]{hTFO}%
  \BibitemOpen
  \bibfield  {author} {\bibinfo {author} {\bibfnamefont {S.-J.}\ \bibnamefont
  {Ahn}}, \bibinfo {author} {\bibfnamefont {J.-H.}\ \bibnamefont {Lee}},
  \bibinfo {author} {\bibfnamefont {H.~M.}\ \bibnamefont {Jang}},\ and\
  \bibinfo {author} {\bibfnamefont {Y.~K.}\ \bibnamefont {Jeong}},\ }\bibfield
  {title} {\bibinfo {title} {Multiferroism in hexagonally stabilized
  {TmFeO}$_3$ thin films below {120\,K}},\ }\href
  {https://doi.org/10.1039/C4TC00461B} {\bibfield  {journal} {\bibinfo
  {journal} {J. Mater. Chem. C}\ }\textbf {\bibinfo {volume} {2}},\ \bibinfo
  {pages} {4521} (\bibinfo {year} {2014})}\BibitemShut {NoStop}%
\bibitem [{\citenamefont {Jin}\ \emph {et~al.}(2019)\citenamefont {Jin},
  \citenamefont {He}, \citenamefont {Zhang}, \citenamefont {Zhang},
  \citenamefont {Wei},\ and\ \citenamefont {Zhong}}]{hTFO2}%
  \BibitemOpen
  \bibfield  {author} {\bibinfo {author} {\bibfnamefont {L.}~\bibnamefont
  {Jin}}, \bibinfo {author} {\bibfnamefont {Y.}~\bibnamefont {He}}, \bibinfo
  {author} {\bibfnamefont {D.}~\bibnamefont {Zhang}}, \bibinfo {author}
  {\bibfnamefont {H.}~\bibnamefont {Zhang}}, \bibinfo {author} {\bibfnamefont
  {M.}~\bibnamefont {Wei}},\ and\ \bibinfo {author} {\bibfnamefont
  {Z.}~\bibnamefont {Zhong}},\ }\bibfield  {title} {\bibinfo {title}
  {Near-ultraviolet photodetector based on hexagonal {TmFeO}$_3$ ferroelectric
  semiconductor thin film with photovoltaic and pyroelectric effects},\ }\href
  {https://doi.org/10.1063/1.5128702} {\bibfield  {journal} {\bibinfo
  {journal} {{APL} Materials}\ }\textbf {\bibinfo {volume} {7}},\ \bibinfo
  {pages} {121105} (\bibinfo {year} {2019})}\BibitemShut {NoStop}%
\bibitem [{\citenamefont {Bossak}\ \emph {et~al.}(2004)\citenamefont {Bossak},
  \citenamefont {Graboy}, \citenamefont {Gorbenko}, \citenamefont {Kaul},
  \citenamefont {Kartavtseva}, \citenamefont {Svetchnikov},\ and\ \citenamefont
  {Zandbergen}}]{TFOthin1}%
  \BibitemOpen
  \bibfield  {author} {\bibinfo {author} {\bibfnamefont {A.~A.}\ \bibnamefont
  {Bossak}}, \bibinfo {author} {\bibfnamefont {I.~E.}\ \bibnamefont {Graboy}},
  \bibinfo {author} {\bibfnamefont {O.~Y.}\ \bibnamefont {Gorbenko}}, \bibinfo
  {author} {\bibfnamefont {A.~R.}\ \bibnamefont {Kaul}}, \bibinfo {author}
  {\bibfnamefont {M.~S.}\ \bibnamefont {Kartavtseva}}, \bibinfo {author}
  {\bibfnamefont {V.~L.}\ \bibnamefont {Svetchnikov}},\ and\ \bibinfo {author}
  {\bibfnamefont {H.~W.}\ \bibnamefont {Zandbergen}},\ }\bibfield  {title}
  {\bibinfo {title} {{XRD} and {HREM} studies of epitaxially stabilized
  hexagonal orthoferrites {RFeO}$_3$ ({R = Eu-Lu})},\ }\href
  {https://doi.org/10.1021/cm0353660} {\bibfield  {journal} {\bibinfo
  {journal} {Chemistry of Materials}\ }\textbf {\bibinfo {volume} {16}},\
  \bibinfo {pages} {1751} (\bibinfo {year} {2004})}\BibitemShut {NoStop}%
\bibitem [{\citenamefont {Akbashev}\ \emph {et~al.}(2011)\citenamefont
  {Akbashev}, \citenamefont {Semisalova}, \citenamefont {Perov},\ and\
  \citenamefont {Kaul}}]{Akbashev2011}%
  \BibitemOpen
  \bibfield  {author} {\bibinfo {author} {\bibfnamefont {A.~R.}\ \bibnamefont
  {Akbashev}}, \bibinfo {author} {\bibfnamefont {A.~S.}\ \bibnamefont
  {Semisalova}}, \bibinfo {author} {\bibfnamefont {N.~S.}\ \bibnamefont
  {Perov}},\ and\ \bibinfo {author} {\bibfnamefont {A.~R.}\ \bibnamefont
  {Kaul}},\ }\bibfield  {title} {\bibinfo {title} {Weak ferromagnetism in
  hexagonal orthoferrites {RFeO}3 (r{\hspace{0.167em}}={\hspace{0.167em}}lu,
  er-tb)},\ }\href {https://doi.org/10.1063/1.3643043} {\bibfield  {journal}
  {\bibinfo  {journal} {Applied Physics Letters}\ }\textbf {\bibinfo {volume}
  {99}},\ \bibinfo {pages} {122502} (\bibinfo {year} {2011})}\BibitemShut
  {NoStop}%
\bibitem [{\citenamefont {Marezio}\ \emph {et~al.}(1970)\citenamefont
  {Marezio}, \citenamefont {Remeika},\ and\ \citenamefont {Dernier}}]{Marezio}%
  \BibitemOpen
  \bibfield  {author} {\bibinfo {author} {\bibfnamefont {M.}~\bibnamefont
  {Marezio}}, \bibinfo {author} {\bibfnamefont {J.~P.}\ \bibnamefont
  {Remeika}},\ and\ \bibinfo {author} {\bibfnamefont {P.~D.}\ \bibnamefont
  {Dernier}},\ }\bibfield  {title} {\bibinfo {title} {{The crystal chemistry of
  the rare earth orthoferrites}},\ }\href
  {https://doi.org/10.1107/S0567740870005319} {\bibfield  {journal} {\bibinfo
  {journal} {Acta Crystallographica Section B}\ }\textbf {\bibinfo {volume}
  {26}},\ \bibinfo {pages} {2008} (\bibinfo {year} {1970})}\BibitemShut
  {NoStop}%
\bibitem [{\citenamefont {Leake}\ \emph {et~al.}(1968)\citenamefont {Leake},
  \citenamefont {Shirane},\ and\ \citenamefont {Remeika}}]{LEAKE1968}%
  \BibitemOpen
  \bibfield  {author} {\bibinfo {author} {\bibfnamefont {J.}~\bibnamefont
  {Leake}}, \bibinfo {author} {\bibfnamefont {G.}~\bibnamefont {Shirane}},\
  and\ \bibinfo {author} {\bibfnamefont {J.}~\bibnamefont {Remeika}},\
  }\bibfield  {title} {\bibinfo {title} {The magnetic structure of thulium
  orthoferrite, {TmFeO}$_3$},\ }\href
  {https://doi.org/10.1016/0038-1098(68)90327-X} {\bibfield  {journal}
  {\bibinfo  {journal} {Solid State Communications}\ }\textbf {\bibinfo
  {volume} {6}},\ \bibinfo {pages} {15 } (\bibinfo {year} {1968})}\BibitemShut
  {NoStop}%
\bibitem [{\citenamefont {Wolfe}\ \emph {et~al.}(1967)\citenamefont {Wolfe},
  \citenamefont {Pierce}, \citenamefont {Haszko},\ and\ \citenamefont
  {Remeika}}]{MagTFO2}%
  \BibitemOpen
  \bibfield  {author} {\bibinfo {author} {\bibfnamefont {R.}~\bibnamefont
  {Wolfe}}, \bibinfo {author} {\bibfnamefont {R.~D.}\ \bibnamefont {Pierce}},
  \bibinfo {author} {\bibfnamefont {S.~E.}\ \bibnamefont {Haszko}},\ and\
  \bibinfo {author} {\bibfnamefont {J.~P.}\ \bibnamefont {Remeika}},\
  }\bibfield  {title} {\bibinfo {title} {Temperature-induced spin reorientation
  in rare earth orthoferrites {\textemdash} hysteresis loop studies},\ }\href
  {https://doi.org/10.1063/1.1755118} {\bibfield  {journal} {\bibinfo
  {journal} {Applied Physics Letters}\ }\textbf {\bibinfo {volume} {11}},\
  \bibinfo {pages} {245} (\bibinfo {year} {1967})}\BibitemShut {NoStop}%
\bibitem [{\citenamefont {White}(1969)}]{White1969}%
  \BibitemOpen
  \bibfield  {author} {\bibinfo {author} {\bibfnamefont {R.~L.}\ \bibnamefont
  {White}},\ }\bibfield  {title} {\bibinfo {title} {Review of recent work on
  the magnetic and spectroscopic properties of the rare-earth orthoferrites},\
  }\href {https://doi.org/10.1063/1.1657530} {\bibfield  {journal} {\bibinfo
  {journal} {Journal of Applied Physics}\ }\textbf {\bibinfo {volume} {40}},\
  \bibinfo {pages} {1061} (\bibinfo {year} {1969})}\BibitemShut {NoStop}%
\bibitem [{\citenamefont {Treves}(1965)}]{MagTFO}%
  \BibitemOpen
  \bibfield  {author} {\bibinfo {author} {\bibfnamefont {D.}~\bibnamefont
  {Treves}},\ }\bibfield  {title} {\bibinfo {title} {Studies on orthoferrites
  at the {Weizmann} institute of science},\ }\href
  {https://doi.org/10.1063/1.1714088} {\bibfield  {journal} {\bibinfo
  {journal} {Journal of Applied Physics}\ }\textbf {\bibinfo {volume} {36}},\
  \bibinfo {pages} {1033} (\bibinfo {year} {1965})}\BibitemShut {NoStop}%
\bibitem [{\citenamefont {Staub}\ \emph {et~al.}(2017)\citenamefont {Staub},
  \citenamefont {Rettig}, \citenamefont {Bothschafter}, \citenamefont
  {Windsor}, \citenamefont {Ramakrishnan}, \citenamefont {Avula}, \citenamefont
  {Dreiser}, \citenamefont {Piamonteze}, \citenamefont {Scagnoli},
  \citenamefont {Mukherjee}, \citenamefont {Niedermayer}, \citenamefont
  {Medarde},\ and\ \citenamefont {Pomjakushina}}]{Staub}%
  \BibitemOpen
  \bibfield  {author} {\bibinfo {author} {\bibfnamefont {U.}~\bibnamefont
  {Staub}}, \bibinfo {author} {\bibfnamefont {L.}~\bibnamefont {Rettig}},
  \bibinfo {author} {\bibfnamefont {E.~M.}\ \bibnamefont {Bothschafter}},
  \bibinfo {author} {\bibfnamefont {Y.~W.}\ \bibnamefont {Windsor}}, \bibinfo
  {author} {\bibfnamefont {M.}~\bibnamefont {Ramakrishnan}}, \bibinfo {author}
  {\bibfnamefont {S.~R.~V.}\ \bibnamefont {Avula}}, \bibinfo {author}
  {\bibfnamefont {J.}~\bibnamefont {Dreiser}}, \bibinfo {author} {\bibfnamefont
  {C.}~\bibnamefont {Piamonteze}}, \bibinfo {author} {\bibfnamefont
  {V.}~\bibnamefont {Scagnoli}}, \bibinfo {author} {\bibfnamefont
  {S.}~\bibnamefont {Mukherjee}}, \bibinfo {author} {\bibfnamefont
  {C.}~\bibnamefont {Niedermayer}}, \bibinfo {author} {\bibfnamefont
  {M.}~\bibnamefont {Medarde}},\ and\ \bibinfo {author} {\bibfnamefont
  {E.}~\bibnamefont {Pomjakushina}},\ }\bibfield  {title} {\bibinfo {title}
  {Interplay of {Fe} and {Tm} moments through the spin-reorientation transition
  in {TmFeO}$_3$},\ }\href {https://doi.org/10.1103/PhysRevB.96.174408}
  {\bibfield  {journal} {\bibinfo  {journal} {Phys. Rev. B}\ }\textbf {\bibinfo
  {volume} {96}},\ \bibinfo {pages} {174408} (\bibinfo {year}
  {2017})}\BibitemShut {NoStop}%
\bibitem [{\citenamefont {Kuroda}\ \emph {et~al.}(2018)\citenamefont {Kuroda},
  \citenamefont {Tanahashi}, \citenamefont {Hajiri}, \citenamefont {Ueda},\
  and\ \citenamefont {Asano}}]{StrainSRT}%
  \BibitemOpen
  \bibfield  {author} {\bibinfo {author} {\bibfnamefont {M.}~\bibnamefont
  {Kuroda}}, \bibinfo {author} {\bibfnamefont {N.}~\bibnamefont {Tanahashi}},
  \bibinfo {author} {\bibfnamefont {T.}~\bibnamefont {Hajiri}}, \bibinfo
  {author} {\bibfnamefont {K.}~\bibnamefont {Ueda}},\ and\ \bibinfo {author}
  {\bibfnamefont {H.}~\bibnamefont {Asano}},\ }\bibfield  {title} {\bibinfo
  {title} {Strain effect on magnetic property of antiferromagnetic insulator
  {SmFeO}$_3$},\ }\href {https://doi.org/10.1063/1.5007332} {\bibfield
  {journal} {\bibinfo  {journal} {AIP Advances}\ }\textbf {\bibinfo {volume}
  {8}},\ \bibinfo {pages} {055814} (\bibinfo {year} {2018})}\BibitemShut
  {NoStop}%
\bibitem [{\citenamefont {Zhou}\ \emph {et~al.}(2020)\citenamefont {Zhou},
  \citenamefont {Marshall}, \citenamefont {Li}, \citenamefont {Li},\ and\
  \citenamefont {He}}]{Zhou2020}%
  \BibitemOpen
  \bibfield  {author} {\bibinfo {author} {\bibfnamefont {J.-S.}\ \bibnamefont
  {Zhou}}, \bibinfo {author} {\bibfnamefont {L.~G.}\ \bibnamefont {Marshall}},
  \bibinfo {author} {\bibfnamefont {Z.-Y.}\ \bibnamefont {Li}}, \bibinfo
  {author} {\bibfnamefont {X.}~\bibnamefont {Li}},\ and\ \bibinfo {author}
  {\bibfnamefont {J.-M.}\ \bibnamefont {He}},\ }\bibfield  {title} {\bibinfo
  {title} {Weak ferromagnetism in perovskite oxides},\ }\href
  {https://doi.org/10.1103/PhysRevB.102.104420} {\bibfield  {journal} {\bibinfo
   {journal} {Phys. Rev. B}\ }\textbf {\bibinfo {volume} {102}},\ \bibinfo
  {pages} {104420} (\bibinfo {year} {2020})}\BibitemShut {NoStop}%
\bibitem [{\citenamefont {Momma}\ and\ \citenamefont {Izumi}(2011)}]{Vesta}%
  \BibitemOpen
  \bibfield  {author} {\bibinfo {author} {\bibfnamefont {K.}~\bibnamefont
  {Momma}}\ and\ \bibinfo {author} {\bibfnamefont {F.}~\bibnamefont {Izumi}},\
  }\bibfield  {title} {\bibinfo {title} {{{\it VESTA3} for three-dimensional
  visualization of crystal, volumetric and morphology data}},\ }\href
  {https://doi.org/10.1107/S0021889811038970} {\bibfield  {journal} {\bibinfo
  {journal} {Journal of Applied Crystallography}\ }\textbf {\bibinfo {volume}
  {44}},\ \bibinfo {pages} {1272} (\bibinfo {year} {2011})}\BibitemShut
  {NoStop}%
\bibitem [{\citenamefont {Nakayama}\ \emph {et~al.}(2013)\citenamefont
  {Nakayama}, \citenamefont {Althammer}, \citenamefont {Chen}, \citenamefont
  {Uchida}, \citenamefont {Kajiwara}, \citenamefont {Kikuchi}, \citenamefont
  {Ohtani}, \citenamefont {Gepr\"ags}, \citenamefont {Opel}, \citenamefont
  {Takahashi}, \citenamefont {Gross}, \citenamefont {Bauer}, \citenamefont
  {Goennenwein},\ and\ \citenamefont {Saitoh}}]{Nakayama2013}%
  \BibitemOpen
  \bibfield  {author} {\bibinfo {author} {\bibfnamefont {H.}~\bibnamefont
  {Nakayama}}, \bibinfo {author} {\bibfnamefont {M.}~\bibnamefont {Althammer}},
  \bibinfo {author} {\bibfnamefont {Y.-T.}\ \bibnamefont {Chen}}, \bibinfo
  {author} {\bibfnamefont {K.}~\bibnamefont {Uchida}}, \bibinfo {author}
  {\bibfnamefont {Y.}~\bibnamefont {Kajiwara}}, \bibinfo {author}
  {\bibfnamefont {D.}~\bibnamefont {Kikuchi}}, \bibinfo {author} {\bibfnamefont
  {T.}~\bibnamefont {Ohtani}}, \bibinfo {author} {\bibfnamefont
  {S.}~\bibnamefont {Gepr\"ags}}, \bibinfo {author} {\bibfnamefont
  {M.}~\bibnamefont {Opel}}, \bibinfo {author} {\bibfnamefont {S.}~\bibnamefont
  {Takahashi}}, \bibinfo {author} {\bibfnamefont {R.}~\bibnamefont {Gross}},
  \bibinfo {author} {\bibfnamefont {G.~E.~W.}\ \bibnamefont {Bauer}}, \bibinfo
  {author} {\bibfnamefont {S.~T.~B.}\ \bibnamefont {Goennenwein}},\ and\
  \bibinfo {author} {\bibfnamefont {E.}~\bibnamefont {Saitoh}},\ }\bibfield
  {title} {\bibinfo {title} {Spin {H}all magnetoresistance induced by a
  nonequilibrium proximity effect},\ }\href
  {https://doi.org/10.1103/PhysRevLett.110.206601} {\bibfield  {journal}
  {\bibinfo  {journal} {Phys. Rev. Lett.}\ }\textbf {\bibinfo {volume} {110}},\
  \bibinfo {pages} {206601} (\bibinfo {year} {2013})}\BibitemShut {NoStop}%
\bibitem [{\citenamefont {Lebrun}\ \emph {et~al.}(2019)\citenamefont {Lebrun},
  \citenamefont {Ross}, \citenamefont {Gomonay}, \citenamefont {Bender},
  \citenamefont {Baldrati}, \citenamefont {Kronast}, \citenamefont
  {Qaiumzadeh}, \citenamefont {Sinova}, \citenamefont {Brataas}, \citenamefont
  {Duine},\ and\ \citenamefont {Kl\"{a}ui}}]{Lebrun2019}%
  \BibitemOpen
  \bibfield  {author} {\bibinfo {author} {\bibfnamefont {R.}~\bibnamefont
  {Lebrun}}, \bibinfo {author} {\bibfnamefont {A.}~\bibnamefont {Ross}},
  \bibinfo {author} {\bibfnamefont {O.}~\bibnamefont {Gomonay}}, \bibinfo
  {author} {\bibfnamefont {S.~A.}\ \bibnamefont {Bender}}, \bibinfo {author}
  {\bibfnamefont {L.}~\bibnamefont {Baldrati}}, \bibinfo {author}
  {\bibfnamefont {F.}~\bibnamefont {Kronast}}, \bibinfo {author} {\bibfnamefont
  {A.}~\bibnamefont {Qaiumzadeh}}, \bibinfo {author} {\bibfnamefont
  {J.}~\bibnamefont {Sinova}}, \bibinfo {author} {\bibfnamefont
  {A.}~\bibnamefont {Brataas}}, \bibinfo {author} {\bibfnamefont {R.~A.}\
  \bibnamefont {Duine}},\ and\ \bibinfo {author} {\bibfnamefont
  {M.}~\bibnamefont {Kl\"{a}ui}},\ }\bibfield  {title} {\bibinfo {title}
  {Anisotropies and magnetic phase transitions in insulating antiferromagnets
  determined by a {Spin-Hall} magnetoresistance probe},\ }\href
  {https://doi.org/10.1038/s42005-019-0150-8} {\bibfield  {journal} {\bibinfo
  {journal} {Communications Physics}\ }\textbf {\bibinfo {volume} {2}},\
  \bibinfo {pages} {50} (\bibinfo {year} {2019})}\BibitemShut {NoStop}%
\bibitem [{\citenamefont {Hajiri}\ \emph {et~al.}(2019)\citenamefont {Hajiri},
  \citenamefont {Baldrati}, \citenamefont {Lebrun}, \citenamefont {Filianina},
  \citenamefont {Ross}, \citenamefont {Tanahashi}, \citenamefont {Kuroda},
  \citenamefont {Gan}, \citenamefont {Mente{\c{s}}}, \citenamefont {Genuzio},
  \citenamefont {Locatelli}, \citenamefont {Asano},\ and\ \citenamefont
  {Kläui}}]{SMRSFO}%
  \BibitemOpen
  \bibfield  {author} {\bibinfo {author} {\bibfnamefont {T.}~\bibnamefont
  {Hajiri}}, \bibinfo {author} {\bibfnamefont {L.}~\bibnamefont {Baldrati}},
  \bibinfo {author} {\bibfnamefont {R.}~\bibnamefont {Lebrun}}, \bibinfo
  {author} {\bibfnamefont {M.}~\bibnamefont {Filianina}}, \bibinfo {author}
  {\bibfnamefont {A.}~\bibnamefont {Ross}}, \bibinfo {author} {\bibfnamefont
  {N.}~\bibnamefont {Tanahashi}}, \bibinfo {author} {\bibfnamefont
  {M.}~\bibnamefont {Kuroda}}, \bibinfo {author} {\bibfnamefont {W.~L.}\
  \bibnamefont {Gan}}, \bibinfo {author} {\bibfnamefont {T.~O.}\ \bibnamefont
  {Mente{\c{s}}}}, \bibinfo {author} {\bibfnamefont {F.}~\bibnamefont
  {Genuzio}}, \bibinfo {author} {\bibfnamefont {A.}~\bibnamefont {Locatelli}},
  \bibinfo {author} {\bibfnamefont {H.}~\bibnamefont {Asano}},\ and\ \bibinfo
  {author} {\bibfnamefont {M.}~\bibnamefont {Kläui}},\ }\bibfield  {title}
  {\bibinfo {title} {Spin structure and spin {Hall} magnetoresistance of
  epitaxial thin films of the insulating non-collinear antiferromagnet
  {SmFeO}$_3$},\ }\href {https://doi.org/10.1088/1361-648x/ab303c} {\bibfield
  {journal} {\bibinfo  {journal} {Journal of Physics: Condensed Matter}\
  }\textbf {\bibinfo {volume} {31}},\ \bibinfo {pages} {445804} (\bibinfo
  {year} {2019})}\BibitemShut {NoStop}%
\bibitem [{\citenamefont {Isasa}\ \emph {et~al.}(2016)\citenamefont {Isasa},
  \citenamefont {V\'elez}, \citenamefont {Sagasta}, \citenamefont
  {Bedoya-Pinto}, \citenamefont {Dix}, \citenamefont {S\'anchez}, \citenamefont
  {Hueso}, \citenamefont {Fontcuberta},\ and\ \citenamefont
  {Casanova}}]{Isasa2016}%
  \BibitemOpen
  \bibfield  {author} {\bibinfo {author} {\bibfnamefont {M.}~\bibnamefont
  {Isasa}}, \bibinfo {author} {\bibfnamefont {S.}~\bibnamefont {V\'elez}},
  \bibinfo {author} {\bibfnamefont {E.}~\bibnamefont {Sagasta}}, \bibinfo
  {author} {\bibfnamefont {A.}~\bibnamefont {Bedoya-Pinto}}, \bibinfo {author}
  {\bibfnamefont {N.}~\bibnamefont {Dix}}, \bibinfo {author} {\bibfnamefont
  {F.}~\bibnamefont {S\'anchez}}, \bibinfo {author} {\bibfnamefont {L.~E.}\
  \bibnamefont {Hueso}}, \bibinfo {author} {\bibfnamefont {J.}~\bibnamefont
  {Fontcuberta}},\ and\ \bibinfo {author} {\bibfnamefont {F.}~\bibnamefont
  {Casanova}},\ }\bibfield  {title} {\bibinfo {title} {Spin {H}all
  magnetoresistance as a probe for surface magnetization in
  {Pt/CoFe}$_2${O}$_4$ bilayers},\ }\href
  {https://doi.org/10.1103/PhysRevApplied.6.034007} {\bibfield  {journal}
  {\bibinfo  {journal} {Phys. Rev. Applied}\ }\textbf {\bibinfo {volume} {6}},\
  \bibinfo {pages} {034007} (\bibinfo {year} {2016})}\BibitemShut {NoStop}%
\bibitem [{\citenamefont {V\'elez}\ \emph {et~al.}(2016)\citenamefont
  {V\'elez}, \citenamefont {Golovach}, \citenamefont {Bedoya-Pinto},
  \citenamefont {Isasa}, \citenamefont {Sagasta}, \citenamefont {Abadia},
  \citenamefont {Rogero}, \citenamefont {Hueso}, \citenamefont {Bergeret},\
  and\ \citenamefont {Casanova}}]{Hanle}%
  \BibitemOpen
  \bibfield  {author} {\bibinfo {author} {\bibfnamefont {S.}~\bibnamefont
  {V\'elez}}, \bibinfo {author} {\bibfnamefont {V.~N.}\ \bibnamefont
  {Golovach}}, \bibinfo {author} {\bibfnamefont {A.}~\bibnamefont
  {Bedoya-Pinto}}, \bibinfo {author} {\bibfnamefont {M.}~\bibnamefont {Isasa}},
  \bibinfo {author} {\bibfnamefont {E.}~\bibnamefont {Sagasta}}, \bibinfo
  {author} {\bibfnamefont {M.}~\bibnamefont {Abadia}}, \bibinfo {author}
  {\bibfnamefont {C.}~\bibnamefont {Rogero}}, \bibinfo {author} {\bibfnamefont
  {L.~E.}\ \bibnamefont {Hueso}}, \bibinfo {author} {\bibfnamefont {F.~S.}\
  \bibnamefont {Bergeret}},\ and\ \bibinfo {author} {\bibfnamefont
  {F.}~\bibnamefont {Casanova}},\ }\bibfield  {title} {\bibinfo {title} {Hanle
  magnetoresistance in thin metal films with strong spin-orbit coupling},\
  }\href {https://doi.org/10.1103/PhysRevLett.116.016603} {\bibfield  {journal}
  {\bibinfo  {journal} {Phys. Rev. Lett.}\ }\textbf {\bibinfo {volume} {116}},\
  \bibinfo {pages} {016603} (\bibinfo {year} {2016})}\BibitemShut {NoStop}%
\bibitem [{\citenamefont {Karplus}\ and\ \citenamefont
  {Luttinger}(1954)}]{Karplus1954}%
  \BibitemOpen
  \bibfield  {author} {\bibinfo {author} {\bibfnamefont {R.}~\bibnamefont
  {Karplus}}\ and\ \bibinfo {author} {\bibfnamefont {J.~M.}\ \bibnamefont
  {Luttinger}},\ }\bibfield  {title} {\bibinfo {title} {Hall effect in
  ferromagnetics},\ }\href {https://doi.org/10.1103/PhysRev.95.1154} {\bibfield
   {journal} {\bibinfo  {journal} {Phys. Rev.}\ }\textbf {\bibinfo {volume}
  {95}},\ \bibinfo {pages} {1154} (\bibinfo {year} {1954})}\BibitemShut
  {NoStop}%
\bibitem [{\citenamefont {Chen}\ \emph {et~al.}(2016)\citenamefont {Chen},
  \citenamefont {Takahashi}, \citenamefont {Nakayama}, \citenamefont
  {Althammer}, \citenamefont {Goennenwein}, \citenamefont {Saitoh},\ and\
  \citenamefont {Bauer}}]{SMR}%
  \BibitemOpen
  \bibfield  {author} {\bibinfo {author} {\bibfnamefont {Y.-T.}\ \bibnamefont
  {Chen}}, \bibinfo {author} {\bibfnamefont {S.}~\bibnamefont {Takahashi}},
  \bibinfo {author} {\bibfnamefont {H.}~\bibnamefont {Nakayama}}, \bibinfo
  {author} {\bibfnamefont {M.}~\bibnamefont {Althammer}}, \bibinfo {author}
  {\bibfnamefont {S.~T.~B.}\ \bibnamefont {Goennenwein}}, \bibinfo {author}
  {\bibfnamefont {E.}~\bibnamefont {Saitoh}},\ and\ \bibinfo {author}
  {\bibfnamefont {G.~E.~W.}\ \bibnamefont {Bauer}},\ }\bibfield  {title}
  {\bibinfo {title} {Theory of spin {Hall} magnetoresistance ({SMR}) and
  related phenomena},\ }\href {https://doi.org/10.1088/0953-8984/28/10/103004}
  {\bibfield  {journal} {\bibinfo  {journal} {Journal of Physics: Condensed
  Matter}\ }\textbf {\bibinfo {volume} {28}},\ \bibinfo {pages} {103004}
  (\bibinfo {year} {2016})}\BibitemShut {NoStop}%
\bibitem [{\citenamefont {Taskin}\ \emph {et~al.}(2017)\citenamefont {Taskin},
  \citenamefont {Legg}, \citenamefont {Yang}, \citenamefont {Sasaki},
  \citenamefont {Kanai}, \citenamefont {Matsumoto}, \citenamefont {Rosch},\
  and\ \citenamefont {Ando}}]{Planar}%
  \BibitemOpen
  \bibfield  {author} {\bibinfo {author} {\bibfnamefont {A.}~\bibnamefont
  {Taskin}}, \bibinfo {author} {\bibfnamefont {H.~F.}\ \bibnamefont {Legg}},
  \bibinfo {author} {\bibfnamefont {F.}~\bibnamefont {Yang}}, \bibinfo {author}
  {\bibfnamefont {S.}~\bibnamefont {Sasaki}}, \bibinfo {author} {\bibfnamefont
  {Y.}~\bibnamefont {Kanai}}, \bibinfo {author} {\bibfnamefont
  {K.}~\bibnamefont {Matsumoto}}, \bibinfo {author} {\bibfnamefont
  {A.}~\bibnamefont {Rosch}},\ and\ \bibinfo {author} {\bibfnamefont
  {Y.}~\bibnamefont {Ando}},\ }\bibfield  {title} {\bibinfo {title} {Planar
  hall effect from the surface of topological insulators},\ }\href
  {https://doi.org/10.1038/s41467-017-01474-8} {\bibfield  {journal} {\bibinfo
  {journal} {Nature communications}\ }\textbf {\bibinfo {volume} {8}},\
  \bibinfo {pages} {1} (\bibinfo {year} {2017})}\BibitemShut {NoStop}%
\bibitem [{\citenamefont {Kimata}\ \emph {et~al.}(2019)\citenamefont {Kimata},
  \citenamefont {Chen}, \citenamefont {Kondou}, \citenamefont {Sugimoto},
  \citenamefont {Muduli}, \citenamefont {Ikhlas}, \citenamefont {Omori},
  \citenamefont {Tomita}, \citenamefont {MacDonald}, \citenamefont
  {Nakatsuji},\ and\ \citenamefont {Otani}}]{Kimata2019}%
  \BibitemOpen
  \bibfield  {author} {\bibinfo {author} {\bibfnamefont {M.}~\bibnamefont
  {Kimata}}, \bibinfo {author} {\bibfnamefont {H.}~\bibnamefont {Chen}},
  \bibinfo {author} {\bibfnamefont {K.}~\bibnamefont {Kondou}}, \bibinfo
  {author} {\bibfnamefont {S.}~\bibnamefont {Sugimoto}}, \bibinfo {author}
  {\bibfnamefont {P.~K.}\ \bibnamefont {Muduli}}, \bibinfo {author}
  {\bibfnamefont {M.}~\bibnamefont {Ikhlas}}, \bibinfo {author} {\bibfnamefont
  {Y.}~\bibnamefont {Omori}}, \bibinfo {author} {\bibfnamefont
  {T.}~\bibnamefont {Tomita}}, \bibinfo {author} {\bibfnamefont {A.~H.}\
  \bibnamefont {MacDonald}}, \bibinfo {author} {\bibfnamefont {S.}~\bibnamefont
  {Nakatsuji}},\ and\ \bibinfo {author} {\bibfnamefont {Y.}~\bibnamefont
  {Otani}},\ }\bibfield  {title} {\bibinfo {title} {Magnetic and
  magnetic~inverse spin {H}all effects in a non-collinear antiferromagnet},\
  }\href {https://doi.org/10.1038/s41586-018-0853-0} {\bibfield  {journal}
  {\bibinfo  {journal} {Nature}\ }\textbf {\bibinfo {volume} {565}},\ \bibinfo
  {pages} {627} (\bibinfo {year} {2019})}\BibitemShut {NoStop}%
\bibitem [{\citenamefont {Binz}\ and\ \citenamefont
  {Vishwanath}(2008)}]{Binz2008}%
  \BibitemOpen
  \bibfield  {author} {\bibinfo {author} {\bibfnamefont {B.}~\bibnamefont
  {Binz}}\ and\ \bibinfo {author} {\bibfnamefont {A.}~\bibnamefont
  {Vishwanath}},\ }\bibfield  {title} {\bibinfo {title} {Chirality induced
  anomalous-hall effect in helical spin crystals},\ }\href
  {https://doi.org/10.1016/j.physb.2007.10.136} {\bibfield  {journal} {\bibinfo
   {journal} {Physica B: Condensed Matter}\ }\textbf {\bibinfo {volume}
  {403}},\ \bibinfo {pages} {1336} (\bibinfo {year} {2008})}\BibitemShut
  {NoStop}%
\bibitem [{\citenamefont {Vlietstra}\ \emph {et~al.}(2013)\citenamefont
  {Vlietstra}, \citenamefont {Shan}, \citenamefont {Castel}, \citenamefont
  {Youssef}, \citenamefont {Bauer},\ and\ \citenamefont {van Wees}}]{SMRYIG}%
  \BibitemOpen
  \bibfield  {author} {\bibinfo {author} {\bibfnamefont {N.}~\bibnamefont
  {Vlietstra}}, \bibinfo {author} {\bibfnamefont {J.}~\bibnamefont {Shan}},
  \bibinfo {author} {\bibfnamefont {V.}~\bibnamefont {Castel}}, \bibinfo
  {author} {\bibfnamefont {J.~B.}\ \bibnamefont {Youssef}}, \bibinfo {author}
  {\bibfnamefont {G.~E.~W.}\ \bibnamefont {Bauer}},\ and\ \bibinfo {author}
  {\bibfnamefont {B.~J.}\ \bibnamefont {van Wees}},\ }\bibfield  {title}
  {\bibinfo {title} {Exchange magnetic field torques in {YIG}/{Pt} bilayers
  observed by the spin-{Hall} magnetoresistance},\ }\href
  {https://doi.org/10.1063/1.4813760} {\bibfield  {journal} {\bibinfo
  {journal} {Applied Physics Letters}\ }\textbf {\bibinfo {volume} {103}},\
  \bibinfo {pages} {032401} (\bibinfo {year} {2013})}\BibitemShut {NoStop}%
\bibitem [{\citenamefont {Chen}\ \emph {et~al.}(2013)\citenamefont {Chen},
  \citenamefont {Takahashi}, \citenamefont {Nakayama}, \citenamefont
  {Althammer}, \citenamefont {Goennenwein}, \citenamefont {Saitoh},\ and\
  \citenamefont {Bauer}}]{SMR1}%
  \BibitemOpen
  \bibfield  {author} {\bibinfo {author} {\bibfnamefont {Y.-T.}\ \bibnamefont
  {Chen}}, \bibinfo {author} {\bibfnamefont {S.}~\bibnamefont {Takahashi}},
  \bibinfo {author} {\bibfnamefont {H.}~\bibnamefont {Nakayama}}, \bibinfo
  {author} {\bibfnamefont {M.}~\bibnamefont {Althammer}}, \bibinfo {author}
  {\bibfnamefont {S.~T.~B.}\ \bibnamefont {Goennenwein}}, \bibinfo {author}
  {\bibfnamefont {E.}~\bibnamefont {Saitoh}},\ and\ \bibinfo {author}
  {\bibfnamefont {G.~E.~W.}\ \bibnamefont {Bauer}},\ }\bibfield  {title}
  {\bibinfo {title} {Theory of spin hall magnetoresistance},\ }\href
  {https://doi.org/10.1103/PhysRevB.87.144411} {\bibfield  {journal} {\bibinfo
  {journal} {Phys. Rev. B}\ }\textbf {\bibinfo {volume} {87}},\ \bibinfo
  {pages} {144411} (\bibinfo {year} {2013})}\BibitemShut {NoStop}%
\bibitem [{\citenamefont {E.A.~Balykina}(1987)}]{jetp1}%
  \BibitemOpen
  \bibfield  {author} {\bibinfo {author} {\bibfnamefont {G.~K.}\ \bibnamefont
  {E.A.~Balykina}, \bibfnamefont {E.A.~Gan'shina}},\ }\bibfield  {title}
  {\bibinfo {title} {Magnetooptic properties of rare-earth orthoferrites in the
  region of spin reorientation transitions},\ }\href@noop {} {\bibfield
  {journal} {\bibinfo  {journal} {Journal of Experimental and Theoretical
  Physics}\ }\textbf {\bibinfo {volume} {66}},\ \bibinfo {pages} {1073}
  (\bibinfo {year} {1987})}\BibitemShut {NoStop}%
\bibitem [{\citenamefont {Gorodetsky}\ \emph {et~al.}(1981)\citenamefont
  {Gorodetsky}, \citenamefont {Shaft},\ and\ \citenamefont
  {Remeika}}]{SurfaceMagnetoelastic}%
  \BibitemOpen
  \bibfield  {author} {\bibinfo {author} {\bibfnamefont {G.}~\bibnamefont
  {Gorodetsky}}, \bibinfo {author} {\bibfnamefont {S.}~\bibnamefont {Shaft}},\
  and\ \bibinfo {author} {\bibfnamefont {J.~P.}\ \bibnamefont {Remeika}},\
  }\bibfield  {title} {\bibinfo {title} {Propagation of surface magnetoelastic
  waves in {TmFeO}$_3$ at the spin reorientation},\ }\href
  {https://doi.org/10.1063/1.328683} {\bibfield  {journal} {\bibinfo  {journal}
  {Journal of Applied Physics}\ }\textbf {\bibinfo {volume} {52}},\ \bibinfo
  {pages} {7353} (\bibinfo {year} {1981})}\BibitemShut {NoStop}%
\bibitem [{\citenamefont {Shang}\ \emph {et~al.}(2016)\citenamefont {Shang},
  \citenamefont {Zhan}, \citenamefont {Yang}, \citenamefont {Zuo},
  \citenamefont {Xie}, \citenamefont {Liu}, \citenamefont {Zhang},
  \citenamefont {Zhang}, \citenamefont {Li}, \citenamefont {Wang},
  \citenamefont {Wu}, \citenamefont {Zhang},\ and\ \citenamefont
  {Li}}]{NioProx}%
  \BibitemOpen
  \bibfield  {author} {\bibinfo {author} {\bibfnamefont {T.}~\bibnamefont
  {Shang}}, \bibinfo {author} {\bibfnamefont {Q.~F.}\ \bibnamefont {Zhan}},
  \bibinfo {author} {\bibfnamefont {H.~L.}\ \bibnamefont {Yang}}, \bibinfo
  {author} {\bibfnamefont {Z.~H.}\ \bibnamefont {Zuo}}, \bibinfo {author}
  {\bibfnamefont {Y.~L.}\ \bibnamefont {Xie}}, \bibinfo {author} {\bibfnamefont
  {L.~P.}\ \bibnamefont {Liu}}, \bibinfo {author} {\bibfnamefont {S.~L.}\
  \bibnamefont {Zhang}}, \bibinfo {author} {\bibfnamefont {Y.}~\bibnamefont
  {Zhang}}, \bibinfo {author} {\bibfnamefont {H.~H.}\ \bibnamefont {Li}},
  \bibinfo {author} {\bibfnamefont {B.~M.}\ \bibnamefont {Wang}}, \bibinfo
  {author} {\bibfnamefont {Y.~H.}\ \bibnamefont {Wu}}, \bibinfo {author}
  {\bibfnamefont {S.}~\bibnamefont {Zhang}},\ and\ \bibinfo {author}
  {\bibfnamefont {R.-W.}\ \bibnamefont {Li}},\ }\bibfield  {title} {\bibinfo
  {title} {Effect of {NiO} inserted layer on spin-{H}all magnetoresistance in
  {Pt}/{NiO}/{YIG} heterostructures},\ }\href
  {https://doi.org/10.1063/1.4959573} {\bibfield  {journal} {\bibinfo
  {journal} {Applied Physics Letters}\ }\textbf {\bibinfo {volume} {109}},\
  \bibinfo {pages} {032410} (\bibinfo {year} {2016})}\BibitemShut {NoStop}%
\bibitem [{\citenamefont {LeCraw}\ \emph {et~al.}(1968)\citenamefont {LeCraw},
  \citenamefont {Wolfe}, \citenamefont {Gyorgy}, \citenamefont {Hagedorn},
  \citenamefont {Hensel},\ and\ \citenamefont {Remeika}}]{LeCraw1968}%
  \BibitemOpen
  \bibfield  {author} {\bibinfo {author} {\bibfnamefont {R.~C.}\ \bibnamefont
  {LeCraw}}, \bibinfo {author} {\bibfnamefont {R.}~\bibnamefont {Wolfe}},
  \bibinfo {author} {\bibfnamefont {E.~M.}\ \bibnamefont {Gyorgy}}, \bibinfo
  {author} {\bibfnamefont {F.~B.}\ \bibnamefont {Hagedorn}}, \bibinfo {author}
  {\bibfnamefont {J.~C.}\ \bibnamefont {Hensel}},\ and\ \bibinfo {author}
  {\bibfnamefont {J.~P.}\ \bibnamefont {Remeika}},\ }\bibfield  {title}
  {\bibinfo {title} {Microwave absorption near the reorientation temperature in
  rare earth orthoferrites},\ }\href {https://doi.org/10.1063/1.1656152}
  {\bibfield  {journal} {\bibinfo  {journal} {Journal of Applied Physics}\
  }\textbf {\bibinfo {volume} {39}},\ \bibinfo {pages} {1019} (\bibinfo {year}
  {1968})}\BibitemShut {NoStop}%
\bibitem [{\citenamefont {Nečas}\ and\ \citenamefont
  {Klapetek}(2012)}]{Gwyddion}%
  \BibitemOpen
  \bibfield  {author} {\bibinfo {author} {\bibfnamefont {D.}~\bibnamefont
  {Nečas}}\ and\ \bibinfo {author} {\bibfnamefont {P.}~\bibnamefont
  {Klapetek}},\ }\bibfield  {title} {\bibinfo {title} {Gwyddion: an open-source
  software for {SPM} data analysis},\ }\href
  {https://doi.org/10.2478/s11534-011-0096-2} {\bibfield  {journal} {\bibinfo
  {journal} {Central European Journal of Physics}\ }\textbf {\bibinfo {volume}
  {10}},\ \bibinfo {pages} {181} (\bibinfo {year} {2012})}\BibitemShut
  {NoStop}%
\bibitem [{\citenamefont {Ross}\ \emph {et~al.}(2019)\citenamefont {Ross},
  \citenamefont {Lebrun}, \citenamefont {Gomonay}, \citenamefont {Grave},
  \citenamefont {Kay}, \citenamefont {Baldrati}, \citenamefont {Becker},
  \citenamefont {Qaiumzadeh}, \citenamefont {Ulloa}, \citenamefont {Jakob},
  \citenamefont {Kronast}, \citenamefont {Sinova}, \citenamefont {Duine},
  \citenamefont {Brataas}, \citenamefont {Rothschild},\ and\ \citenamefont
  {Kl\"{a}ui}}]{transport}%
  \BibitemOpen
  \bibfield  {author} {\bibinfo {author} {\bibfnamefont {A.}~\bibnamefont
  {Ross}}, \bibinfo {author} {\bibfnamefont {R.}~\bibnamefont {Lebrun}},
  \bibinfo {author} {\bibfnamefont {O.}~\bibnamefont {Gomonay}}, \bibinfo
  {author} {\bibfnamefont {D.~A.}\ \bibnamefont {Grave}}, \bibinfo {author}
  {\bibfnamefont {A.}~\bibnamefont {Kay}}, \bibinfo {author} {\bibfnamefont
  {L.}~\bibnamefont {Baldrati}}, \bibinfo {author} {\bibfnamefont
  {S.}~\bibnamefont {Becker}}, \bibinfo {author} {\bibfnamefont
  {A.}~\bibnamefont {Qaiumzadeh}}, \bibinfo {author} {\bibfnamefont
  {C.}~\bibnamefont {Ulloa}}, \bibinfo {author} {\bibfnamefont
  {G.}~\bibnamefont {Jakob}}, \bibinfo {author} {\bibfnamefont
  {F.}~\bibnamefont {Kronast}}, \bibinfo {author} {\bibfnamefont
  {J.}~\bibnamefont {Sinova}}, \bibinfo {author} {\bibfnamefont
  {R.}~\bibnamefont {Duine}}, \bibinfo {author} {\bibfnamefont
  {A.}~\bibnamefont {Brataas}}, \bibinfo {author} {\bibfnamefont
  {A.}~\bibnamefont {Rothschild}},\ and\ \bibinfo {author} {\bibfnamefont
  {M.}~\bibnamefont {Kl\"{a}ui}},\ }\bibfield  {title} {\bibinfo {title}
  {Propagation length of antiferromagnetic magnons governed by domain
  configurations},\ }\href {https://doi.org/10.1021/acs.nanolett.9b03837}
  {\bibfield  {journal} {\bibinfo  {journal} {Nano Letters}\ }\textbf {\bibinfo
  {volume} {20}},\ \bibinfo {pages} {306} (\bibinfo {year} {2019})}\BibitemShut
  {NoStop}%
\bibitem [{\citenamefont {Horng}\ \emph {et~al.}(2004)\citenamefont {Horng},
  \citenamefont {Chern}, \citenamefont {Chen}, \citenamefont {Kang},\ and\
  \citenamefont {Lee}}]{CFO1}%
  \BibitemOpen
  \bibfield  {author} {\bibinfo {author} {\bibfnamefont {L.}~\bibnamefont
  {Horng}}, \bibinfo {author} {\bibfnamefont {G.}~\bibnamefont {Chern}},
  \bibinfo {author} {\bibfnamefont {M.}~\bibnamefont {Chen}}, \bibinfo {author}
  {\bibfnamefont {P.}~\bibnamefont {Kang}},\ and\ \bibinfo {author}
  {\bibfnamefont {D.}~\bibnamefont {Lee}},\ }\bibfield  {title} {\bibinfo
  {title} {Magnetic anisotropic properties in {Fe}$_3${O}$_4$ and
  {CoFe}$_2${O}$_4$ ferrite epitaxy thin films},\ }\href
  {https://doi.org/10.1016/j.jmmm.2003.09.005} {\bibfield  {journal} {\bibinfo
  {journal} {Journal of Magnetism and Magnetic Materials}\ }\textbf {\bibinfo
  {volume} {270}},\ \bibinfo {pages} {389 } (\bibinfo {year}
  {2004})}\BibitemShut {NoStop}%
\bibitem [{\citenamefont {Rigato}\ \emph {et~al.}(2009)\citenamefont {Rigato},
  \citenamefont {Geshev}, \citenamefont {Skumryev},\ and\ \citenamefont
  {Fontcuberta}}]{CFO2}%
  \BibitemOpen
  \bibfield  {author} {\bibinfo {author} {\bibfnamefont {F.}~\bibnamefont
  {Rigato}}, \bibinfo {author} {\bibfnamefont {J.}~\bibnamefont {Geshev}},
  \bibinfo {author} {\bibfnamefont {V.}~\bibnamefont {Skumryev}},\ and\
  \bibinfo {author} {\bibfnamefont {J.}~\bibnamefont {Fontcuberta}},\
  }\bibfield  {title} {\bibinfo {title} {The magnetization of epitaxial
  nanometric {CoFe}$_2${O}$_4$ (001) layers},\ }\href
  {https://doi.org/10.1063/1.3267873} {\bibfield  {journal} {\bibinfo
  {journal} {Journal of Applied Physics}\ }\textbf {\bibinfo {volume} {106}},\
  \bibinfo {pages} {113924} (\bibinfo {year} {2009})}\BibitemShut {NoStop}%
\bibitem [{\citenamefont {Zhang}\ \emph {et~al.}(2008)\citenamefont {Zhang},
  \citenamefont {Deng}, \citenamefont {Ma}, \citenamefont {Lin},\ and\
  \citenamefont {Nan}}]{CFO3}%
  \BibitemOpen
  \bibfield  {author} {\bibinfo {author} {\bibfnamefont {Y.}~\bibnamefont
  {Zhang}}, \bibinfo {author} {\bibfnamefont {C.}~\bibnamefont {Deng}},
  \bibinfo {author} {\bibfnamefont {J.}~\bibnamefont {Ma}}, \bibinfo {author}
  {\bibfnamefont {Y.}~\bibnamefont {Lin}},\ and\ \bibinfo {author}
  {\bibfnamefont {C.-W.}\ \bibnamefont {Nan}},\ }\bibfield  {title} {\bibinfo
  {title} {Enhancement in magnetoelectric response in
  {CoFe}$_2${O}$_4$–{BaTiO}$_3$ heterostructure},\ }\href
  {https://doi.org/10.1063/1.2841048} {\bibfield  {journal} {\bibinfo
  {journal} {Applied Physics Letters}\ }\textbf {\bibinfo {volume} {92}},\
  \bibinfo {pages} {062911} (\bibinfo {year} {2008})}\BibitemShut {NoStop}%
\bibitem [{\citenamefont {Rodewald}\ \emph {et~al.}(2019)\citenamefont
  {Rodewald}, \citenamefont {Thien}, \citenamefont {Pohlmann}, \citenamefont
  {Hoppe}, \citenamefont {Timmer}, \citenamefont {Bertram}, \citenamefont
  {Kuepper},\ and\ \citenamefont {Wollschl\"ager}}]{CFO4}%
  \BibitemOpen
  \bibfield  {author} {\bibinfo {author} {\bibfnamefont {J.}~\bibnamefont
  {Rodewald}}, \bibinfo {author} {\bibfnamefont {J.}~\bibnamefont {Thien}},
  \bibinfo {author} {\bibfnamefont {T.}~\bibnamefont {Pohlmann}}, \bibinfo
  {author} {\bibfnamefont {M.}~\bibnamefont {Hoppe}}, \bibinfo {author}
  {\bibfnamefont {F.}~\bibnamefont {Timmer}}, \bibinfo {author} {\bibfnamefont
  {F.}~\bibnamefont {Bertram}}, \bibinfo {author} {\bibfnamefont
  {K.}~\bibnamefont {Kuepper}},\ and\ \bibinfo {author} {\bibfnamefont
  {J.}~\bibnamefont {Wollschl\"ager}},\ }\bibfield  {title} {\bibinfo {title}
  {Formation of ultrathin cobalt ferrite films by interdiffusion of
  {Fe}$_3${O}$_4$/{CoO} bilayers},\ }\href
  {https://doi.org/10.1103/PhysRevB.100.155418} {\bibfield  {journal} {\bibinfo
   {journal} {Phys. Rev. B}\ }\textbf {\bibinfo {volume} {100}},\ \bibinfo
  {pages} {155418} (\bibinfo {year} {2019})}\BibitemShut {NoStop}%
\bibitem [{\citenamefont {Schmool}\ \emph
  {et~al.}(1999{\natexlab{a}})\citenamefont {Schmool}, \citenamefont {Keller},
  \citenamefont {Guyot}, \citenamefont {Krishnan},\ and\ \citenamefont
  {Tessier}}]{Garnet}%
  \BibitemOpen
  \bibfield  {author} {\bibinfo {author} {\bibfnamefont {D.~S.}\ \bibnamefont
  {Schmool}}, \bibinfo {author} {\bibfnamefont {N.}~\bibnamefont {Keller}},
  \bibinfo {author} {\bibfnamefont {M.}~\bibnamefont {Guyot}}, \bibinfo
  {author} {\bibfnamefont {R.}~\bibnamefont {Krishnan}},\ and\ \bibinfo
  {author} {\bibfnamefont {M.}~\bibnamefont {Tessier}},\ }\bibfield  {title}
  {\bibinfo {title} {Magnetic and magneto-optic properties of orthoferrite thin
  films grown by pulsed-laser deposition},\ }\href
  {https://doi.org/10.1063/1.371583} {\bibfield  {journal} {\bibinfo  {journal}
  {Journal of Applied Physics}\ }\textbf {\bibinfo {volume} {86}},\ \bibinfo
  {pages} {5712} (\bibinfo {year} {1999}{\natexlab{a}})}\BibitemShut {NoStop}%
\bibitem [{\citenamefont {Schmool}\ \emph
  {et~al.}(1999{\natexlab{b}})\citenamefont {Schmool}, \citenamefont {Keller},
  \citenamefont {Guyot}, \citenamefont {Krishnan},\ and\ \citenamefont
  {Tessier}}]{Magnetite}%
  \BibitemOpen
  \bibfield  {author} {\bibinfo {author} {\bibfnamefont {D.}~\bibnamefont
  {Schmool}}, \bibinfo {author} {\bibfnamefont {N.}~\bibnamefont {Keller}},
  \bibinfo {author} {\bibfnamefont {M.}~\bibnamefont {Guyot}}, \bibinfo
  {author} {\bibfnamefont {R.}~\bibnamefont {Krishnan}},\ and\ \bibinfo
  {author} {\bibfnamefont {M.}~\bibnamefont {Tessier}},\ }\bibfield  {title}
  {\bibinfo {title} {Evidence of very high coercive fields in orthoferrite
  phases of {PLD} grown thin films},\ }\href
  {https://doi.org/10.1016/s0304-8853(99)00102-x} {\bibfield  {journal}
  {\bibinfo  {journal} {Journal of Magnetism and Magnetic Materials}\ }\textbf
  {\bibinfo {volume} {195}},\ \bibinfo {pages} {291} (\bibinfo {year}
  {1999}{\natexlab{b}})}\BibitemShut {NoStop}%
\bibitem [{\citenamefont {Pauthenet}(1958)}]{Pauthenet1958}%
  \BibitemOpen
  \bibfield  {author} {\bibinfo {author} {\bibfnamefont {R.}~\bibnamefont
  {Pauthenet}},\ }\bibfield  {title} {\bibinfo {title} {Spontaneous
  magnetization of some garnet ferrites and the aluminum substituted garnet
  ferrites},\ }\href {https://doi.org/10.1063/1.1723094} {\bibfield  {journal}
  {\bibinfo  {journal} {Journal of Applied Physics}\ }\textbf {\bibinfo
  {volume} {29}},\ \bibinfo {pages} {253} (\bibinfo {year} {1958})}\BibitemShut
  {NoStop}%
\bibitem [{\citenamefont {\"{O}zden \"{O}zdemir}\ \emph
  {et~al.}(2002)\citenamefont {\"{O}zden \"{O}zdemir}, \citenamefont {Dunlop},\
  and\ \citenamefont {Moskowitz}}]{zdemir2002}%
  \BibitemOpen
  \bibfield  {author} {\bibinfo {author} {\bibnamefont {\"{O}zden
  \"{O}zdemir}}, \bibinfo {author} {\bibfnamefont {D.~J.}\ \bibnamefont
  {Dunlop}},\ and\ \bibinfo {author} {\bibfnamefont {B.~M.}\ \bibnamefont
  {Moskowitz}},\ }\bibfield  {title} {\bibinfo {title} {Changes in remanence,
  coercivity and domain state at low temperature in magnetite},\ }\href
  {https://doi.org/10.1016/s0012-821x(01)00562-3} {\bibfield  {journal}
  {\bibinfo  {journal} {Earth and Planetary Science Letters}\ }\textbf
  {\bibinfo {volume} {194}},\ \bibinfo {pages} {343} (\bibinfo {year}
  {2002})}\BibitemShut {NoStop}%
\bibitem [{\citenamefont {Gondek}\ \emph {et~al.}(2010)\citenamefont {Gondek},
  \citenamefont {Kaczorowski},\ and\ \citenamefont {Szytu{\l}a}}]{Gondek2010}%
  \BibitemOpen
  \bibfield  {author} {\bibinfo {author} {\bibfnamefont {{\L}.}~\bibnamefont
  {Gondek}}, \bibinfo {author} {\bibfnamefont {D.}~\bibnamefont
  {Kaczorowski}},\ and\ \bibinfo {author} {\bibfnamefont {A.}~\bibnamefont
  {Szytu{\l}a}},\ }\bibfield  {title} {\bibinfo {title} {Low temperature
  studies on magnetic properties of {Tm}$_2${O}$_3$},\ }\href
  {https://doi.org/10.1016/j.ssc.2009.11.032} {\bibfield  {journal} {\bibinfo
  {journal} {Solid State Communications}\ }\textbf {\bibinfo {volume} {150}},\
  \bibinfo {pages} {368} (\bibinfo {year} {2010})}\BibitemShut {NoStop}%
\bibitem [{\citenamefont {Barbier}\ \emph {et~al.}(2005)\citenamefont
  {Barbier}, \citenamefont {Belkhou}, \citenamefont {Ohresser}, \citenamefont
  {Gautier-Soyer}, \citenamefont {Bezencenet}, \citenamefont {Mulazzi},
  \citenamefont {Guittet},\ and\ \citenamefont {Moussy}}]{Barbier2005}%
  \BibitemOpen
  \bibfield  {author} {\bibinfo {author} {\bibfnamefont {A.}~\bibnamefont
  {Barbier}}, \bibinfo {author} {\bibfnamefont {R.}~\bibnamefont {Belkhou}},
  \bibinfo {author} {\bibfnamefont {P.}~\bibnamefont {Ohresser}}, \bibinfo
  {author} {\bibfnamefont {M.}~\bibnamefont {Gautier-Soyer}}, \bibinfo {author}
  {\bibfnamefont {O.}~\bibnamefont {Bezencenet}}, \bibinfo {author}
  {\bibfnamefont {M.}~\bibnamefont {Mulazzi}}, \bibinfo {author} {\bibfnamefont
  {M.-J.}\ \bibnamefont {Guittet}},\ and\ \bibinfo {author} {\bibfnamefont
  {J.-B.}\ \bibnamefont {Moussy}},\ }\bibfield  {title} {\bibinfo {title}
  {Electronic and crystalline structure, morphology, and magnetism of
  nanometric {Fe}$_2${O}$_3$ layers deposited on {Pt}(111) by
  atomic-oxygen-assisted molecular beam epitaxy},\ }\href
  {https://doi.org/10.1103/PhysRevB.72.245423} {\bibfield  {journal} {\bibinfo
  {journal} {Phys. Rev. B}\ }\textbf {\bibinfo {volume} {72}},\ \bibinfo
  {pages} {245423} (\bibinfo {year} {2005})}\BibitemShut {NoStop}%
\end{thebibliography}%

\end{document}